\newif\ifjnr
\jnrtrue

\ifjnr
  \documentclass[a4paper,10pt]{article}
\else
  \documentclass[onecolumn]{elsart3p}
\fi

\usepackage{amssymb}
\usepackage[fleqn]{amsmath}
\usepackage{mathdots}
\usepackage[square,sort&compress,numbers]{natbib}

\usepackage{subfigure}
\usepackage{ifpdf}
\ifpdf
  \usepackage[pdftex]{hyperref}
  \usepackage[pdftex]{graphicx}
\else
  \usepackage[ps2pdf,colorlinks=true]{hyperref}
  \usepackage[dvips]{graphicx}
\fi

\ifjnr
  \usepackage{fancyhdr}
  \usepackage{geometry}  
  \usepackage{multicol}  
  \usepackage{times}

  \renewcommand{\headrulewidth}{0pt}
  \makeatletter
  
  \newenvironment{figurehere}
  {\def\@captype{figure}}
  {}
  \makeatother

\fi

\begin{document}


\newcommand{\eqn}[1]{(#1)}
\newcommand{\Eqn}[1]{(#1)}
\newcommand{\tbl}[1]{Table~#1}
\newcommand{\Tbl}[1]{Table~#1}
\newcommand{\fig}[1]{Fig.~#1}
\newcommand{\Fig}[1]{Fig.~#1}
\newcommand{\sectn}[1]{Sec.~#1}
\newcommand{\Sectn}[1]{Sec.~#1}
\newcommand{\chap}[1]{Chapter~#1}
\newcommand{\Chap}[1]{Chapter~#1}
\newcommand{\appn}[1]{Appendix~#1}
\newcommand{\Appn}[1]{Appendix~#1}
\newcommand{\lemmaref}[1]{Lemma~#1}

\newcommand{\etal}{\mbox{\it et al.}}
\newcommand{\eg}{\mbox{\it e.g.}}
\newcommand{\ie}{\mbox{\it i.e.}}
\newcommand{\etc}{\mbox{\it etc.}}
\newcommand{\cf}{\mbox{\it cf.}}


\newcommand{\ghz}{{GHz}}
\newcommand{\joule}{{J}}
\newcommand{\watt}{{W}}
\newcommand{\metre}{m}
\newcommand{\nanometre}{nm}
\newcommand{\steradian}{sr}
\newcommand{\kelvin}{{K}}
\newcommand{\arcmin}{\ensuremath{{}^\prime}}
\newcommand{\degrees}{\ensuremath{{^\circ}}}


\newcommand{\fft}{{FFT}}
\newcommand{\dft}{{DFT}}

\newcommand{\frw}{{FRW}}
\newcommand{\frwtext}{{\fbox{Friedmann}-Robertson-Walker}}

\newcommand{\cmb}{{CMB}}
\newcommand{\cmbtext}{{cosmic microwave background}}
\newcommand{\Cmbtext}{{Cosmic microwave background}}

\newcommand{\wmap}{{WMAP}}
\newcommand{\wmaptext}{{Wilkinson Microwave Anisotropy Probe}}
\newcommand{\ilc}{{ILC1}}
\newcommand{\ilcthree}{{ILC3}}
\newcommand{\ilctext}{{internal linear combination}}

\newcommand{\cobe}{\mbox{COBE-DMR}}
\newcommand{\cobetext}{Cosmic Background Explorer-Differential Microwave Radiometer}

\newcommand{\bianchi}{{Bianchi}}
\newcommand{\bianchiviih}{{Bianchi VII{$_{\lowercase{h}}$}}}
\newcommand{\wmapbianchi}{{\wmap1$-$Bianchi}}

\newcommand{\isw}{{ISW}}
\newcommand{\iswtext}{{integrated Sachs-Wolfe}}
\newcommand{\Iswtext}{{Integrated Sachs-Wolfe}}

\newcommand{\cswt}{{CSWT}}
\newcommand{\cswti}{\mbox{CSWT$_{\rm I}$}}
\newcommand{\cswtii}{\mbox{CSWT$_{\rm II}$}}
\newcommand{\cswttext}{continuous spherical wavelet transform}
\newcommand{\Cswttext}{Continuous spherical wavelet transform}

\newcommand{\lss}{{LSS}}
\newcommand{\lsstext}{large scale structure}
\newcommand{\Lsstext}{Large scale structure}

\newcommand{\nvss}{{NVSS}}
\newcommand{\nvsstext}{{NRAO VLA Sky Survey}}

\newcommand{\sdss}{{SDSS}}
\newcommand{\sdsstext}{Sloan Digital Sky Survey}

\newcommand{\twomass}{{2MASS}}
\newcommand{\twomasstext}{Two Micron All Sky Survey}

\newcommand{\lcdm}{\ensuremath{\Lambda}{CDM}}
\newcommand{\lcdmtext}{\ensuremath{\Lambda} Cold Dark Matter}

\newcommand{\smhwtext}{{spherical Mexican hat wavelet}}
\newcommand{\smhw}{{SMHW}}
\newcommand{\semhw}{{SEMHW}}
\newcommand{\smw}{{SMW}}
\newcommand{\sbwtext}{{spherical butterfly wavelet}}
\newcommand{\sbw}{{SBW}}
\newcommand{\mexhat}{Mexican hat}
\newcommand{\Mexhat}{Mexican hat}
\newcommand{\morlet}{real Morlet}
\newcommand{\Morlet}{Real Morlet}

\newcommand{\healpix}{{\tt HEALPix}}
\newcommand{\healpixtext}{{Hierarchical Equal Area isoLatitude Pixelisation of the sphere}}
\newcommand{\cmbfast}{\mbox{\tt CMBFAST}}
\newcommand{\camb}{\mbox{\tt CAMB}}
\newcommand{\cosmomc}{\mbox{\tt COSMOMC}}

\newcommand{\yawtbtext}{{Yet Another Wavelet Toolbox}}
\newcommand{\yawtb}{{\tt YAWTb}}

\newcommand{\stwofil}{{\tt S2FIL}}
\newcommand{\fastcswt}{{\tt FastCSWT}}
\newcommand{\comb}{{\tt COMB}}
\newcommand{\bianchicode}{{\tt Bianchi}}
\newcommand{\bianchicodetwo}{{\tt Bianchi2}}

\newcommand{\lambdaarch}{{LAMBDA}}
\newcommand{\lambdaarchtext}{{Legacy Archive for Microwave Background Data Analysis}}

\newcommand{\kpzero}{{Kp0}}
\newcommand{\tohmap}{{TOH}}

\newcommand{\fwhm}{{FWHM}}
\newcommand{\snr}{{\rm SNR}}

\newcommand{\mf}{{MF}}
\newcommand{\saf}{{SAF}}


\newcommand{\spotloc}{\mbox{\ensuremath{(l,b)=(209^\circ,-57^\circ)}}}

\newcommand{\f}{\ensuremath{f}}
\newcommand{\flm}{\ensuremath{\shc{\f}{\el}{\m}}}
\newcommand{\flmm}{\ensuremath{\shc{\f}{\el,}{-\m}}}
\newcommand{\ft}{\ensuremath{\tilde{\f}}}
\newcommand{\ftlm}{\ensuremath{\shc{\ft}{\el}{\m}}}
\newcommand{\fte}{\ensuremath{\ft^{\rm e}}}
\newcommand{\ftelm}{\ensuremath{\shc{\ft}{\el}{\m}^{\rm e}}}
\newcommand{\fto}{\ensuremath{\ft^{\rm o}}}
\newcommand{\ftolm}{\ensuremath{\shc{\ft}{\el}{\m}^{\rm o}}}

\newcommand{\fs}{\ensuremath{{{}_\spin f}}}
\newcommand{\fsm}{\ensuremath{{{}_{-\spin} f}}}
\newcommand{\fslm}{\ensuremath{\shc{\fs}{\el}{\m}}}
\newcommand{\fslmm}{\ensuremath{\shc{\fs}{\el,}{-\m}}}
\newcommand{\fst}{\ensuremath{{}_\spin \tilde{\f}}}
\newcommand{\fstlm}{\ensuremath{\shc{\fst}{\el}{\m}}}
\newcommand{\fste}{\ensuremath{\fst^{\rm e}}}
\newcommand{\fstelm}{\ensuremath{\shc{\fst}{\el}{\m}^{\rm e}}}
\newcommand{\fsto}{\ensuremath{\fst^{\rm o}}}
\newcommand{\fstolm}{\ensuremath{\shc{\fst}{\el}{\m}^{\rm o}}}

\newcommand{\F}{\ensuremath{F}}
\newcommand{\Ft}{\ensuremath{\tilde{F}}}
\newcommand{\Ffourier}{\ensuremath{\F_{\m\m\p}}}
\newcommand{\Ftfourier}{\ensuremath{\Ft_{\m\m\p}}}
\newcommand{\Ffouriere}{\ensuremath{\Ftfourier^{\rm e}}}
\newcommand{\Ffouriero}{\ensuremath{\Ftfourier^{\rm o}}}

\newcommand{\Fs}{\ensuremath{{}_\spin F}}
\newcommand{\Fst}{\ensuremath{{}_\spin \tilde{\F}}}
\newcommand{\Fsfourier}{\ensuremath{\Fs_{\m\m\p}}}
\newcommand{\Fstfourier}{\ensuremath{\Fst_{\m\m\p}}}
\newcommand{\Fsfouriere}{\ensuremath{\Fstfourier^{\rm e}}}
\newcommand{\Fsfouriero}{\ensuremath{\Fstfourier^{\rm o}}}

\newcommand{\ffts}{\ensuremath{\mathcal{F}}}
\newcommand{\iffts}{\ensuremath{\mathcal{F}^{-1}}}
\newcommand{\fftshift}{\ensuremath{\mathcal{S}}}
\newcommand{\ifftshift}{\ensuremath{\mathcal{S}^{-1}}}

\newcommand{\saai}{\ensuremath{t}}
\newcommand{\sabi}{\ensuremath{p}}
\newcommand{\sais}{\ensuremath{\sabi,\saai}}

\newcommand{\mz}{\ensuremath{{\m_{0}}}}

\newcommand{\spcend}{\ensuremath{\:}}
\newcommand{\img}{\ensuremath{{i}}}
\newcommand{\cconj}{\ensuremath{\ast}} 
\newcommand{\reals}{\ensuremath{\mathbb{R}}}
\newcommand{\realsnn}{\ensuremath{\mathbb{R^+}}}
\newcommand{\realsnz}{\ensuremath{\mathbb{R}^{+}_{\ast}}}
\newcommand{\integers}{\ensuremath{\mathbb{Z}}}
\newcommand{\naturals}{\ensuremath{\mathbb{N}}}
\newcommand{\ltwo}{\ensuremath{\mathrm{L}^2}}
\newcommand{\sphere}{\ensuremath{{\mathrm{S}^2}}}
\newcommand{\sothree}{\ensuremath{{\mathrm{SO}(3)}}}
\newcommand{\torus}{\ensuremath{{\mathrm{T}^2}}}
\newcommand{\vect}[1]{\ensuremath{\mbox{\boldmath ${#1}$}}}
\newcommand{\vectsm}[1]{\ensuremath{\mbox{\boldmath \footnotesize ${#1}$}}}
\newcommand{\xvect}{\vect{x}}
\newcommand{\kvect}{\vect{k}}
\newcommand{\zerovect}{\ensuremath{\mathbf{0}}}
\newcommand{\knaut}{\ensuremath{k_0}}
\newcommand{\nsigma}{\ensuremath{N_\sigma}}
\newcommand{\zreal}{{\ensuremath{\rm{Re}}}}
\newcommand{\zimag}{{\ensuremath{\rm{Im}}}}
\newcommand{\dotprod}{\ensuremath{\cdot}}
\newcommand{\likelihood}{\ensuremath{\mathcal{L}}}
\newcommand{\opnexpv}{\ensuremath{\left \langle}}
\newcommand{\clsexpv}{\ensuremath{\right \rangle}}

\newcommand{\dx}{\ensuremath{\mathrm{\,d}}}
\newcommand{\dmu}[1]{\ensuremath{\dx \Omega(#1)}}
\newcommand{\dmun}{\ensuremath{\dx \Omega}}
\newcommand{\deul}[1]{\ensuremath{\dx \varrho(#1)}}
\newcommand{\deuln}{\ensuremath{\dx \varrho}}
\newcommand{\ddr}{\ensuremath{\frac{\partial}{\partial\scale}}}


\newcommand{\clight}{\ensuremath{c}}
\newcommand{\ctime}{\ensuremath{\tau}}
\newcommand{\ptime}{\ensuremath{t}}
\newcommand{\pdist}{\ensuremath{d}}
\newcommand{\rvel}{\ensuremath{v}}
\newcommand{\hub}{\ensuremath{H}}
\newcommand{\hubsmall}{\ensuremath{h}}
\newcommand{\scfac}{\ensuremath{R}}
\newcommand{\scfacd}{\ensuremath{\dot{\scfac}}}
\newcommand{\scfacdd}{\ensuremath{\ddot{\scfac}}}
\newcommand{\kcurv}{\ensuremath{k}}
\newcommand{\cconst}{\ensuremath{\Lambda}}
\newcommand{\gconst}{\ensuremath{G}}
\newcommand{\den}{\ensuremath{\rho}}
\newcommand{\denmat}{\ensuremath{\rho_{\rm m}}}
\newcommand{\denrad}{\ensuremath{\rho_{\rm r}}}
\newcommand{\dencrit}{\ensuremath{\rho_{\rm c}}}
\newcommand{\Den}{\ensuremath{\Omega}}
\newcommand{\Dentot}{\ensuremath{\Den_{\rm total}}}
\newcommand{\Denmat}{\ensuremath{\Den_{\rm m}}}
\newcommand{\Denrad}{\ensuremath{\Den_{\rm r}}}
\newcommand{\Denlambda}{\ensuremath{\Den_{\Lambda}}}
\newcommand{\pres}{\ensuremath{p}}
\newcommand{\w}{\ensuremath{w}}
\newcommand{\z}{\ensuremath{z}}
\newcommand{\zrec}{\ensuremath{\z_{\rm rec}}}

\newcommand{\dtemp}{\ensuremath{\Delta T}}
\newcommand{\temp}{\ensuremath{T_0}}
\newcommand{\rdtemp}{\ensuremath{\frac{\dtemp}{\temp}}}
\newcommand{\gravpotent}{\ensuremath{\delta\Phi}}
\newcommand{\cl}{\ensuremath{C_\el}}
\newcommand{\clest}{\ensuremath{\widehat{C}_\el}}

\newcommand{\sa}{\ensuremath{\omega}}
\newcommand{\saa}{\ensuremath{\theta}}
\newcommand{\sab}{\ensuremath{\varphi}}
\newcommand{\sas}{\ensuremath{\saa, \sab}}
\newcommand{\eul}{\ensuremath{\mathbf{\rho}}}
\newcommand{\euls}{\ensuremath{\eula, \eulb, \eulc}}
\newcommand{\eula}{\ensuremath{\alpha}}
\newcommand{\eulb}{\ensuremath{\beta}}
\newcommand{\eulc}{\ensuremath{\gamma}}
\newcommand{\el}{\ensuremath{\ell}}
\newcommand{\m}{\ensuremath{m}}
\newcommand{\n}{\ensuremath{n}}
\newcommand{\elm}{\ensuremath{{\el\m}}}
\newcommand{\elp}{\ensuremath{{\el\p}}}
\newcommand{\mmax}{\ensuremath{\m_{\rm max}}}
\newcommand{\elmax}{\ensuremath{{L}}}
\newcommand{\p}{\ensuremath{^\prime}}
\newcommand{\pp}{\ensuremath{^{\prime\prime}}}
\newcommand{\scale}{\ensuremath{R}}
\newcommand{\scalenaut}{\ensuremath{R_0}}
\newcommand{\scalea}{\ensuremath{a}}
\newcommand{\scaleb}{\ensuremath{b}}
\newcommand{\scaleab}{\ensuremath{{\scalea,\scaleb}}}
\newcommand{\rad}{\ensuremath{r}}
\newcommand{\cmbtemp}{\ensuremath{T}}
\newcommand{\pixw}{\ensuremath{p_\el}}
\newcommand{\bm}[1]{\ensuremath{b_\el^{#1}}}

\newcommand{\kron}[2]{\ensuremath{\delta_{{#1}{#2}}}}
\newcommand{\kronsp}[2]{\ensuremath{\delta_{{#1},{#2}}}}
\newcommand{\poly}{\ensuremath{\phi}}
\newcommand{\polyw}{\ensuremath{w}}
\renewcommand{\exp}[1]{\ensuremath{{\rm e}^{#1}}}
\newcommand{\shfarg}[3]{\ensuremath{Y_{#1#2}({#3})}}
\newcommand{\shfargc}[3]{\ensuremath{Y_{#1#2}^\cconj({#3})}}
\newcommand{\shfargsp}[3]{\ensuremath{Y_{{#1},{#2}}({#3})}}

\newcommand{\sshfarg}[4]{\ensuremath{{{}_{#4} Y_{#1#2}({#3})}}}
\newcommand{\sshfargc}[4]{\ensuremath{{{}_{#4} Y_{#1#2}^\cconj({#3})}}}
\newcommand{\sshfargsp}[4]{\ensuremath{{{}_{#4} Y_{{#1},{#2}}({#3})}}}

\newcommand{\shf}[2]{\ensuremath{Y_{#1#2}}}
\newcommand{\shfc}[2]{\ensuremath{Y_{#1#2}^\cconj}}
\newcommand{\shc}[3]{\ensuremath{{#1}_{{#2}{#3}}}}
\newcommand{\shcc}[3]{\ensuremath{{#1}_{{#2}{#3}}^\cconj}}
\newcommand{\shcsp}[3]{\ensuremath{{#1}_{{#2},{#3}}}}
\newcommand{\shccsp}[3]{\ensuremath{{#1}_{{#2},{#3}}^\cconj}}

\newcommand{\sshf}[3]{\ensuremath{{{}_{#3} Y_{#1#2}}}}
\newcommand{\sshc}[4]{\ensuremath{{}_{#4} {#1}_{{#2}{#3}}}}
\newcommand{\sshcc}[4]{\ensuremath{{}_{#4} {#1}_{{#2}{#3}}^\cconj}}
\newcommand{\sshcsp}[4]{\ensuremath{{}_{#4} {#1}_{{#2},{#3}}}}
\newcommand{\sshccsp}[4]{\ensuremath{{}_{#4} {#1}_{{#2},{#3}}^\cconj}}

\newcommand{\spin}{\ensuremath{s}}
\newcommand{\spinup}{\ensuremath{\eth}}
\newcommand{\spindown}{\ensuremath{\bar{\eth}}}

\newcommand{\leg}[2]{\ensuremath{P_{{#1}}({#2})}}
\newcommand{\aleg}[3]{\ensuremath{P_{#1}^{#2}({#3})}}
\newcommand{\legc}[2]{\ensuremath{{#1}_{#2}}}
\newcommand{\daleg}[3]{\ensuremath{P_{#1}^{#2\prime}({#3})}}
\newcommand{\sbessel}[1]{\ensuremath{j_{#1}}}
\newcommand{\bessel}[2]{\ensuremath{J_{#1}({#2})}}
\newcommand{\jacobi}[4]{\ensuremath{P_{#1}^{(#2,#3)}({#4})}}
\newcommand{\jacobia}{\ensuremath{a}}
\newcommand{\jacobib}{\ensuremath{b}}
\newcommand{\gammafun}{\ensuremath{\Gamma}}
\newcommand{\binomial}[2]{\ensuremath{\left( \begin{array}{c} {#1} \\ {#2} \end{array} \right)}}

\newcommand{\dmatbig}{\ensuremath{D}}
\newcommand{\Dlmn}{\ensuremath{ \dmatbig_{\m\n}^{\el} }}
\newcommand{\Dlmnc}{\ensuremath{ \dmatbig_{\m\n}^{\el\cconj} }}
\newcommand{\Dlmz}{\ensuremath{ \dmatbig_{\m0}^{\el} }}
\newcommand{\Dlmzc}{\ensuremath{ \dmatbig_{\m0}^{\el\cconj} }}
\newcommand{\Dlmnp}{\ensuremath{ \dmatbig_{\m\n}^{\el}(\eul) }}
\newcommand{\Dlmnpc}{\ensuremath{ \dmatbig_{\m\n}^{\el\cconj}(\eul) }}
\newcommand{\Dlmnabg}{\ensuremath{ \dmatbig_{\m\n}^{\el}(\euls) }}
\newcommand{\Dlmnabgc}{\ensuremath{ \dmatbig_{\m\n}^{\el\cconj}(\euls) }}
\newcommand{\dmatsmall}{\ensuremath{d}}
\newcommand{\dlmn}{\ensuremath{ \dmatsmall_{\m\n}^{\el} }}
\newcommand{\dlmnc}{\ensuremath{ \dmatsmall_{\m\n}^{\el\cconj} }}
\newcommand{\dlnm}{\ensuremath{ \dmatsmall_{\n\m}^{\el} }}
\newcommand{\dlmz}{\ensuremath{ \dmatsmall_{\m0}^{\el} }}
\newcommand{\dlmzb}{\ensuremath{ \dmatsmall_{\m0}^{\el}(\eulb) }}
\newcommand{\dlmzbc}{\ensuremath{ \dmatsmall_{\m0}^{\el\cconj}(\eulb) }}
\newcommand{\dlmzc}{\ensuremath{ \dmatsmall_{\m0}^{\el\cconj} }}
\newcommand{\dlmnb}{\ensuremath{ \dmatsmall_{\m\n}^{\el}(\eulb) }}
\newcommand{\dlmnbc}{\ensuremath{ \dmatsmall_{\m\n}^{\el\cconj}(\eulb) }}
\newcommand{\dlnmb}{\ensuremath{ \dmatsmall_{\n\m}^{\el}(\eulb) }}

\newcommand{\dlms}{\ensuremath{ \dmatsmall_{\m\spin}^{\el} }}

\newcommand{\dlmnhalfpi}{\ensuremath{ \dmatsmall_{\m\n}^{\el}(\pi/2) }}
\newcommand{\dlmzhalfpi}{\ensuremath{ \dmatsmall_{\m0}^{\el}(\pi/2) }}
\newcommand{\dlmshalfpi}{\ensuremath{ \dmatsmall_{\m0}^{\el}(\pi/2) }}
\newcommand{\dlmmnnhalfpi}{\ensuremath{ \dmatsmall_{-\m,-\n}^{\el}(\pi/2) }}

\newcommand{\dlmmzhalfpi}{\ensuremath{ \dmatsmall_{-\m,0}^{\el}(\pi/2) }}
\newcommand{\dlmmnhalfpi}{\ensuremath{ \dmatsmall_{-\m,\n}^{\el}(\pi/2)
}}

\newcommand{\dlmnarg}[3]{\ensuremath{ \dmatsmall_{{#2},{#3}}^{#1} }}

\newcommand{\dil}{\ensuremath{\mathcal{D}}}
\newcommand{\dilsmall}{\ensuremath{d}}
\newcommand{\spo}{\ensuremath{\Pi}}
\newcommand{\rot}{\ensuremath{\mathcal{R}}}

\newcommand{\sky}{\ensuremath{f}}
\newcommand{\skywav}{\ensuremath{{\mathcal{W}_\wav^\sky}}}
\newcommand{\skywavni}{\ensuremath{{\mathcal{W}}}}
\newcommand{\skywavi}{\ensuremath{{\widehat{\mathcal{W}}_\wavi^\sky}}}
\newcommand{\skyLwavii}{\ensuremath{{\widetilde{\mathcal{W}}_\wavm^{\Lopii\sky}}}}
\newcommand{\skywavii}{\ensuremath{{\widetilde{\mathcal{W}}_\wavm^\sky}}}
\newcommand{\wav}{\ensuremath{\psi}}
\newcommand{\wavi}{\ensuremath{\Phi}}
\newcommand{\wavii}{\ensuremath{\Psi}}
\newcommand{\wavm}{\ensuremath{\Upsilon}}
\newcommand{\wavmc}{\ensuremath{\Upsilon^\cconj}}
\newcommand{\admissC}{\ensuremath{C}}
\newcommand{\admissCi}{\ensuremath{\widehat{\admissC}_\wavi}}
\newcommand{\admissCli}{\ensuremath{\widehat{\admissC}_\wavi^\el}}
\newcommand{\admissCii}{\ensuremath{\widetilde{\admissC}_\wavm}}
\newcommand{\admissClii}{\ensuremath{\widetilde{\admissC}_\wavm^\el}}
\newcommand{\Lopi}{\ensuremath{\widehat{L}_\wavi}}
\newcommand{\Lopii}{\ensuremath{\widetilde{L}_\wavm}}
\newcommand{\cocycle}{\ensuremath{\lambda}}
\newcommand{\effsize}{\ensuremath{\xi}}
\newcommand{\eccen}{\ensuremath{\epsilon}}
\newcommand{\sigx}{\ensuremath{\sigma_x}}
\newcommand{\sigy}{\ensuremath{\sigma_y}}

\newcommand{\elmfact}{\ensuremath{\frac{(\el-\m)!}{(\el+\m)!}}}
\newcommand{\elpifrac}{\ensuremath{\frac{2\el+1}{4\pi}}}
\newcommand{\elpifracinv}{\ensuremath{\frac{4\pi}{2\el+1}}}
\newcommand{\sumlm}{\ensuremath{\sum_{\el=0}^{\infty} \sum_{\m=-\el}^\el}}
\newcommand{\sumlmn}{\ensuremath{\sum_{\el=0}^{\infty} \sum_{\m=-\el}^\el} \sum_{\n=-\el}^\el}
\newcommand{\suml}{\ensuremath{\sum_{\el=0}^{\infty}}}
\newcommand{\summ}{\ensuremath{\sum_{\m=-\el}^\el}}
\newcommand{\sumn}{\ensuremath{\sum_{\n=-\el}^\el}}
\newcommand{\sumulmn}{\ensuremath{\sum_{\el\m\n}}}
\newcommand{\sumlmp}{{\ensuremath{\sum_{\el\p =0}^{\infty} \: \sum_{\m\p=-\el\p }^{\el\p } \:}}}
\newcommand{\sumllm}{{\ensuremath{\sum_{\el =0}^{\infty} \: \sum_{\el\p =0}^{\infty} \: \sum_{\m\p=-\el }^{\el } \:}}}
\newcommand{\nlm}{\ensuremath{\sqrt{\frac{2\el+1}{4\pi}\frac{(\el-\m)!}{(\el+\m)!}}}}
\newcommand{\nlpmp}{\ensuremath{\sqrt{\frac{2\el\p+1}{4\pi}\frac{(\el\p-\m\p)!}{(\el\p+\m\p)!}}}}
\newcommand{\nlpm}{\ensuremath{\sqrt{\frac{2\el\p+1}{4\pi}\frac{(\el\p-\m)!}{(\el\p+\m)!}}}}

\newcommand{\cswtfftterm}{\ensuremath{T}}
\newcommand{\grideula}{\ensuremath{\mathcal{E}_1}}
\newcommand{\grideulb}{\ensuremath{\mathcal{E}_2}}
\newcommand{\gridhpix}{\ensuremath{\mathcal{H}}}
\newcommand{\gridecp}{\ensuremath{\mathcal{C}}}
\newcommand{\ia}{\ensuremath{{n_\eula}}}
\newcommand{\ib}{\ensuremath{{n_\eulb}}}
\newcommand{\ig}{\ensuremath{{n_\eulc}}}
\newcommand{\ik}{\ensuremath{{k}}}
\newcommand{\ith}{\ensuremath{{n_\saa}}}
\newcommand{\iph}{\ensuremath{{n_\sab}}}
\newcommand{\ipix}{\ensuremath{{p}}}
\newcommand{\na}{\ensuremath{{N_\eula}}}
\newcommand{\nb}{\ensuremath{{N_\eulb}}}
\newcommand{\ngm}{\ensuremath{{N_\eulc}}}
\newcommand{\nth}{\ensuremath{{N_\saa}}}
\newcommand{\nph}{\ensuremath{{N_\sab}}}
\newcommand{\N}{\ensuremath{{N}}}
\newcommand{\nside}{\ensuremath{{N_{\rm{side}}}}}
\newcommand{\npix}{\ensuremath{{N_{\rm{pix}}}}}
\newcommand{\weight}{\ensuremath{w}}
\newcommand{\order}{\ensuremath{\mathcal{O}}}

\newcommand{\mean}{\ensuremath{\mu}}
\newcommand{\skewness}{\ensuremath{\zeta}}
\newcommand{\kurtosis}{\ensuremath{\kappa}}
\newcommand{\neff}{\ensuremath{N_{\rm eff}}}
\newcommand{\num}{\ensuremath{N}}
\newcommand{\nstd}{\ensuremath{\num_\sigma}}
\newcommand{\ndev}{\ensuremath{\num_{\rm dev}}}
\newcommand{\nstat}{\ensuremath{\num_{\rm stat}}}
\newcommand{\conflevel}{\ensuremath{\rm SL}}
\newcommand{\cov}{\ensuremath{\mathbf{C}}}
\newcommand{\tstat}{\ensuremath{\tau}}

\newcommand{\bx}{\ensuremath{x}}
\newcommand{\bhand}{\ensuremath{\kappa}}
\newcommand{\bh}{\ensuremath{h}}
\newcommand{\ze}{\ensuremath{\zrec}}
\newcommand{\bshear}{\ensuremath{\left(\frac{\sigma}{H}\right)_0}}
\newcommand{\bvort}{\ensuremath{\left(\frac{\omega}{H}\right)_0}}
\newcommand{\ba}{\ensuremath{A}}
\newcommand{\bi}[2]{\ensuremath{I^{#1}_{#2}}}
\newcommand{\thetaph}{\ensuremath{\theta_0}}
\newcommand{\phiph}{\ensuremath{\phi_0}}
\newcommand{\thetaob}{\ensuremath{\theta_{\rm ob}}}
\newcommand{\phiob}{\ensuremath{\phi_{\rm ob}}}
\newcommand{\thetaalm}{\ensuremath{\theta}}
\newcommand{\phialm}{\ensuremath{\phi}}
\newcommand{\bs}{\ensuremath{s}}
\newcommand{\bt}{\ensuremath{\tau}}
\newcommand{\bps}{\ensuremath{\psi}}
\newcommand{\bcone}{\ensuremath{C_1}}
\newcommand{\bctwo}{\ensuremath{C_2}}
\newcommand{\bcthree}{\ensuremath{C_3}}
\newcommand{\alm}{\ensuremath{a}}
\newcommand{\almi}{\ensuremath{\shc{\alm}{\el}{\m}}}
\newcommand{\almpi}{\ensuremath{\shc{\alm}{\el}{\m\p}}}
\newcommand{\almitilde}{\ensuremath{\shc{\tilde{\alm}}{\el}{\m}}}

\newcommand{\amptmpl}{\ensuremath{\lambda}}
\newcommand{\bparam}{\ensuremath{\Theta_{\rm B}}}
\newcommand{\cosmoparam}{\ensuremath{\Theta_{\rm C}}}
\newcommand{\fitdata}{\ensuremath{\vect{d}}}
\newcommand{\fitmodel}{\ensuremath{\vect{m}}}
\newcommand{\fitmodelsel}{\ensuremath{M}}
\newcommand{\fitamp}{\ensuremath{\lambda}}
\newcommand{\fitampest}{\ensuremath{\widehat{\lambda}}}
\newcommand{\fittmpl}{\ensuremath{\vect{t}}}
\newcommand{\fitnoise}{\ensuremath{\vect{n}}}
\newcommand{\fitcov}{\ensuremath{\mathbf{M}}}
\newcommand{\fitchisqd}{\ensuremath{\chi^2}}
\newcommand{\fitdataalm}{\ensuremath{\shc{d}{\el}{\m}}}
\newcommand{\fittmplalm}{\ensuremath{\shc{t}{\el}{\m}}}
\newcommand{\fitdataalz}{\ensuremath{\shc{d}{\el}{0}}}
\newcommand{\fittmplalz}{\ensuremath{\shc{t}{\el}{0}}}

\newcommand{\wcov}{\ensuremath{X_\wav}}
\newcommand{\wcovest}{\ensuremath{\widehat{X}_\wav}}
\newcommand{\wcovvect}{\ensuremath{\mathbf{X}_\wav}}
\newcommand{\wcovestvect}{\ensuremath{\widehat{\mathbf{X}}_\wav}}
\newcommand{\covweight}{\ensuremath{\nu}}
\newcommand{\covsubterm}{\ensuremath{G^\el}}
\newcommand{\clnt}{\ensuremath{C_\el^{\rm NT, obs}}}
\newcommand{\clnttheo}{\ensuremath{C_\el^{\rm NT}}}
\newcommand{\clnntheo}{\ensuremath{C_\el^{\rm NN}}}
\newcommand{\cltttheo}{\ensuremath{C_\el^{\rm TT}}}
\newcommand{\nd}{\ensuremath{N}}
\newcommand{\tp}{\ensuremath{T}}
\newcommand{\ndlab}{\ensuremath{{\rm N}}}
\newcommand{\tplab}{\ensuremath{{\rm T}}}
\newcommand{\dndz}{\ensuremath{\frac{\dx N}{\dx z}}}
\newcommand{\dgdz}{\ensuremath{\frac{\dx g}{\dx z}}}
\newcommand{\eqdec}{\ensuremath{\delta}}
\newcommand{\chisqd}{\ensuremath{\chi^2}}
\newcommand{\baypar}{\ensuremath{\Theta}}
\newcommand{\lhood}{\ensuremath{\mathcal{L}}}

\newcommand{\fil}{\ensuremath{\varphi}}
\newcommand{\filcoeff}{\ensuremath{w}}
\newcommand{\lagrnmult}{\ensuremath{\mu}}
\newcommand{\lagrn}{\ensuremath{\mathcal{L}}}
\newcommand{\filvara}{\ensuremath{a}}
\newcommand{\filvarb}{\ensuremath{b}}
\newcommand{\filvarc}{\ensuremath{c}}
\newcommand{\filvardenom}{\ensuremath{\Delta}}
\newcommand{\pnorm}{\ensuremath{p}}
\newcommand{\pnormsep}{\ensuremath{|}}
\newcommand{\pnormtext}{\ensuremath{\pnorm}-norm}
\newcommand{\pnormtextit}{\ensuremath{\rm{\pnorm}}-norm}
\newcommand{\scalepnorm}{\ensuremath{{\scale\pnormsep\pnorm}}}
\newcommand{\scalenautpnorm}{\ensuremath{{\scalenaut\pnormsep\pnorm}}}
\newcommand{\za}[2]{\ensuremath{\shc{a}{#1}{#2}}}
\newcommand{\zb}[2]{\ensuremath{\shc{b}{#1}{#2}}}
\newcommand{\zc}[2]{\ensuremath{\shc{c}{#1}{#2}}}
\newcommand{\zd}[2]{\ensuremath{\shc{d}{#1}{#2}}}
\newcommand{\zcsp}[2]{\ensuremath{\shcsp{c}{#1}{#2}}}
\newcommand{\zdsp}[2]{\ensuremath{\shcsp{d}{#1}{#2}}}
\newcommand{\detect}{\ensuremath{\Gamma}}
\newcommand{\beam}{\ensuremath{b}}
\newcommand{\noise}{\ensuremath{n}}
\newcommand{\instnoise}{\ensuremath{r}}
\newcommand{\obsig}{\ensuremath{y}}
\newcommand{\conv}{\ensuremath{\star}}
\newcommand{\amp}{\ensuremath{A}}
\newcommand{\tmpl}{\ensuremath{\tau}}
\newcommand{\noisecl}{\ensuremath{C}}
\newcommand{\amphat}{\ensuremath{\hat{A}}}
\newcommand{\eulshat}{\ensuremath{\hat{\alpha},\hat{\beta},\hat{\gamma}}}


\ifjnr
  \newgeometry{columnsep=10mm,top=13mm,bottom=80mm,left=15mm,right=15mm}
  \setlength{\footskip}{67mm}
  \fancyfootoffset{-15mm}

  \noindent
  \hspace*{15mm}
  \begin{minipage}[center]{150mm}
\else
  \begin{frontmatter}
\fi



\ifjnr\else
  \title{Fast, exact (but unstable)\\spin spherical harmonic transforms}
  \author{J. D. McEwen}
  \ead{mcewen@mrao.cam.ac.uk}
  \address{Astrophysics Group, Cavendish Laboratory,\\
    University of Cambridge, Cambridge CB3 0HE, UK}
\fi

\ifjnr

  \includegraphics[height=18mm]{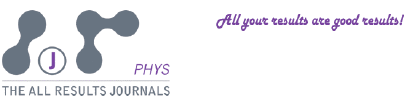}
  \hfill
  \emph{Issue 1, Volume 1, {\bf \emph{2011}}, 4-18}
  \vspace*{5mm}

  \begin{center}
    {\Large \bf Fast, exact (but unstable) spin spherical harmonic transforms}\\*[6mm]

    {\large Jason D. McEwen\\
    Astrophysics Group, Cavendish Laboratory, University of Cambridge,
    UK\\*[0.5mm]
    Institute of Electrical Engineering, Ecole Polytechnique
    F{\'e}d{\'e}rale de Lausanne (EPFL),
    Switzerland}\\*[10mm]

  \end{center}

  {\bf \large Abstract:}
\else
\begin{abstract}
  \fi In many applications data are measured or defined on a spherical
  manifold; spherical harmonic transforms are then required to access
  the frequency content of the data.  We derive\ algorithms to perform forward
  and inverse spin spherical harmonic transforms for functions of
  arbitrary spin number.  These algorithms involve recasting the spin
  transform on the two-sphere \sphere\ as a Fourier transform on the
  two-torus \torus.  Fast Fourier transforms are then used to compute
  Fourier coefficients, which are related to spherical harmonic
  coefficients through a linear transform.  By recasting the problem
  as a Fourier transform on the torus we appeal to the usual Shannon
  sampling theorem to develop spherical harmonic transforms that are
  theoretically exact for band-limited functions, thereby providing an
  alternative sampling theorem on the sphere.  The computational
  complexity of our forward and inverse spin spherical harmonic
  transforms scale as $\order(\elmax^3)$ for any arbitrary spin
  number, where $\elmax$ is the harmonic band-limit of the spin
  function on the sphere.  Numerical experiments are performed and
  unfortunately the forward transform is found to be unstable for
  band-limits above $\elmax\simeq32$.  The instability is due to the
  poorly conditioned linear system relating Fourier and spherical
  harmonic coefficients.  The inverse transform is expected to be
  stable, although it is not possible to verify this hypothesis.
  \ifjnr \\*[1mm] \else
\end{abstract}
\fi

\ifjnr
  {\bf \large Keywords:}
  \newcommand{\sep}{;\,}
\else
  \begin{keyword}
\fi
\emph{spherical harmonics\sep spherical harmonic transform\sep two-sphere\sep algorithms}
\ifjnr
 \\*[7mm]
 \end{minipage}
\else
  \end{keyword}
  \end{frontmatter}
\fi

\ifjnr
  \begin{multicols}{2}
\fi

\section{Introduction}

Data are defined inherently on the sphere in many practical
applications, including
computer graphics (\eg\ \cite{ramamoorthi:2004}), 
\mbox{planetary} science (\eg\ \cite{turcotte:1981,wieczorek:2006,wieczorek:1998}), 
\mbox{solar} physics (\eg\ \cite{stenflo:1986,makarov:2001}),
geophysics (\eg\ \cite{simons:2006,swenson:2002,whaler:1994}),
quantum chemistry (\eg\ \cite{choi:1999,ritchie:1999})
and 
astrophysics 
(\eg\ \cite{bennett:1996,bennett:2003a,hinshaw:2006,hinshaw:2008}),
for example.
In a number of these applications it is insightful to compute the
spherical harmonic representation of the data.  For example,
observations of the anisotropies of the \cmbtext\ (\cmb), that are
inherently made on the celestial sphere, contain a wealth of
information about models of the early Universe.  The anisotropies of
the \cmb\ are seeded by primordial perturbations, which, in linear
perturbation theory, evolve independently for each Fourier mode.
Solutions of models of the early Universe are therefore solved more
conveniently in Fourier space and manifest themselves in the form of
the angular power spectrum of \cmb\ anisotropies.  Cosmologists
compute the angular power spectrum of observations of the \cmb,
through a spherical harmonic transform, to make contact with theory
and to estimate parameters of cosmological models (\eg\
\cite{dunkley:2008}).  Recent and upcoming observations of \cmb\
anisotropies made on the celestial sphere are of considerable size,
with maps containing approximately three \ifjnr\hfill\fi \cite{hinshaw:2008}
\ifjnr\hfill\fi and \ifjnr\hfill\fi fifty \ifjnr\hfill\fi
\cite{planck:bluebook} \ifjnr\hfill\fi million \ifjnr\hfill\fi pixels
\ifjnr\hfill\fi respectively. %
\ifjnr
  \end{multicols}
  \newgeometry{columnsep=10mm,top=22mm,bottom=23mm,left=15mm,right=15mm}
  \fancyhf{}
  \fancyhead[R]{\em All Res. J. Phys., {\bf \em 2011}, 1, 4-18}
  \setlength{\headheight}{7mm}
  \setlength{\headsep}{7mm}
  \setlength{\footskip}{8mm}
  \renewcommand{\headrulewidth}{0pt}
  \fancyfoot[R]{\thepage}
  \fancyfootoffset{0mm}
  \begin{multicols}{2}
  \noindent
\fi
Consequently, the ability to perform a fast spherical harmonic
transform for data of considerable size is essential in assisting
cosmologists to understand our Universe.

A number of works have addressed the problem of computing a spherical
harmonic transform efficiently.  We first survey the related
literature for computing the scalar transform on the sphere, before
discussing approaches for computing the harmonic transform of spin
functions on the sphere.
By performing a simple separation of variables the spherical harmonic
transform may be rewritten as a Fourier transform and an associated
Legendre transform.  This already reduces the complexity of
computation from $\order(\elmax^4)$ to $\order(\elmax^3)$, where
\elmax\ is the harmonic band-limit defining the maximum
frequency content of the function considered (defined mathematically in
  \sectn{\ref{sec:algorithms_harmonic}}).  If this
separation of variable is performed first, then to reduce the
complexity of the spherical harmonic transform further reduces to the
problem of performing a fast associated Legendre transform.

First attempts to compute a scalar spherical harmonic through a fast
Legendre transform were performed by Orszag \cite{orszag:1986} and
were based on a Wentzel-Kramers-Brillouin (WKB) approximation.
Alternative approaches using the fast multipole method (FMM)
\cite{beatson:1997} have been considered by Alpert \& Rokhin
\cite{alpert:1991} and by Suda \& Takani \cite{suda:2002}.  All of
these methods are necessarily approximate, for any choice of
pixelisation of the sphere, and the complexity of these algorithms
scales linearly with the desired accuracy.

Other approaches based on the separation of variables have been
developed for particular pixelisations of the sphere, such as
HEALPix\footnote{\url{http://healpix.jpl.nasa.gov/}\\*[-2mm]}
\cite{gorski:2005},
IGLOO\footnote{\url{http://www.cita.utoronto.ca/~crittend/pixel.html}\\*[-2mm]}
\cite{crittenden:1998} and
GLESP\footnote{\url{http://www.glesp.nbi.dk/}\\*[-2mm]}
\cite{doroshkevich:2005}.
In all of these cases the resulting algorithms are of complexity
$\order(\elmax^3)$ and for HEALPix and IGLOO pixelisations only approximate
quadrature rules exist.
Due to the approximate quadrature the spherical harmonic transform
algorithms of HEALPix and IGLOO are not theoretically exact; a forward
transform followed by an inverse transform will not recover the
original signal perfectly, even with exact-precision arithmetic.  
By contrast GLESP does exhibit an exact quadrature rule due to its
adoption of Gauss-Legendre quadrature.
All of these pixelisation schemes and their associated
harmonic transform algorithms are of sufficient accuracy for many
practical purposes and have been of considerable use in the analysis
of \cmb\ data.

A sampling theorem on the sphere was first developed by Driscoll \&
Healy \cite{driscoll:1994}, providing a framework for a spherical
harmonic transform that is theoretically exact.  Moreover, Driscoll \&
Healy also present a divide-and-conquer approach to computing a fast
associated Legendre transform in the cosine basis, using a fast
Fourier transform (\fft) to project onto this basis.  The resulting
algorithm is exact in exact-precision arithmetic and has computational
complexity $\order(\elmax^2 \log {}_2^2\elmax)$, but is known to
suffer from stability problems \cite{healy:2003,kostelec:2008}.  Potts
\cite{potts:1998} adapted the Driscoll \& Healy approach to use a fast
cosine transform (as suggested previously \cite{driscoll:1994})
to reduce computation time, but not complexity, and introduced some
stabilisation methods.  Healy \etal\ \cite{healy:2003} readdressed the
work of Driscoll \& Healy, reformulating the sampling theorem on the
sphere and developing some variants of the original Driscoll \& Healy
algorithm.  A fast cosine transform is used here and a number of
alternative algorithms are presented, including the so-called
semi-naive, simple split and hybrid algorithms:  the semi-naive
algorithm is a more stable variant of the original (non-divide-and-conquer) Driscoll \& Healy
algorithm with complexity $\order(\elmax^3)$; the simple-split
algorithm is a simpler and more stable divide-and-conquer approach
than the original unstable divide-and-conquer Driscoll \& Healy
algorithm but with a complexity
of \mbox{$\order(\elmax^{5/2} \log {}_2^{1/2} \elmax)$}; the hybrid
algorithm uses both of these two algorithms and splits the problem
between them.  The hybrid algorithm appears to achieve a good
compromise between stability and efficiency.  However, the overall
complexity of this algorithm is not clear since it depends on the
split between the semi-naive and simple split algorithms and on some
user specified parameters.  Implementations of these algorithms
\cite{healy:2003} are available for
download.\footnote{\url{http://www.cs.dartmouth.edu/~geelong/sphere/}}
Due to the divide-and-conquer approach first proposed by Driscoll \&
Healy for the fast associated Legendre transform, these algorithms
\cite{driscoll:1994,potts:1998,healy:2003} are all restricted to
harmonic band-limits that are a power of two.

An alternative algorithm is developed by Mohlenkamp
\cite{mohlenkamp:1997,mohlenkamp:1999}, that is again based on a
separation of variables and a fast associated Legendre transform.
This method is based on the exact sampling theorem on the sphere of
Driscoll \& Healy and a compressed representation of the associated
Legendre functions using WKB frequency estimates.  Rigorous bounds
are derived for the number of coefficients required to achieve a given
accuracy.  The method is again necessarily approximate, but the
approximation can be controlled independently of complexity.  The
complexity of this algorithm is $\order(\elmax^2 \log {}_2^2 \elmax)$
and in practice very good accuracy and stability is achieved.
However, the algorithm is also restricted to band-limits that are a
power of two.

Very little attention has been paid to fast algorithms to perform the
spherical harmonic transform of spin functions.  Spin functions on the
sphere also arise in many applications.  For example, the linear
polarisation of the \cmb\ anisotropies is described by the Stokes
parameters $Q$ and $U$, where the quantity $Q \pm \img U$ is a spin
$\pm2$ function on the sphere \cite{zaldarriaga:1997}.  The reason
little attention has been paid explicitly to the transform of spin
functions may be because spin lowering and raising operators can be
used to relate spin spherical harmonic transforms to the scalar
transform.  This is done explicitly and in a stable manner by Wiaux
\etal\ \cite{wiaux:2005b} for the spin $\pm2$ case.  Scalar spherical
harmonic algorithms developed by Healy \etal\ \cite{healy:2003} are then applied to
yield a theoretically exact spin $\pm2$ algorithm with the same
complexity as the Healy \etal\ algorithm that is adopted.  In general,
this type of approach may be used to compute the spin transform for
arbitrary spin, however the complexity of the resulting spin
algorithm will be scaled by the spin number.

In this work we derive a general algorithm to perform a spin spherical
harmonic transform for functions of arbitrary spin number.  The
complexity of the algorithm is independent of spin number.
Furthermore, any harmonic band-limit may be considered and not only
powers of two.  Our spin spherical harmonic transform algorithm
essentially involves recasting the spin transform on the sphere as a
Fourier transform on the torus.  \fft s are then used to compute
Fourier coefficients, which are related to the spherical harmonic
coefficients of the function considered through a linear transform.
By recasting the problem as a Fourier transform on the torus we appeal
to the usual Shannon sampling theorem \cite{shannon:1949} to develop
spherical harmonic transforms that are theoretically exact for
band-limited functions, thereby providing an alternative sampling
theorem on the sphere.  The remainder of this paper is organised as
follows.  The mathematical background is presented in
\sectn{\ref{sec:preliminaries}}.  In \sectn{\ref{sec:algorithms}} we
derive our spin transform algorithms and discuss a number of
subtleties of the algorithms and practical considerations.  In
\sectn{\ref{sec:experiments}} we examine the memory and computation
complexity of the algorithms and evaluate their stability with
numerical experiments.  Concluding remarks are made in
\sectn{\ref{sec:conclusions}}.

\section{Mathematical background}
\label{sec:preliminaries}

Before deriving our fast algorithms it is necessary to outline some
mathematical preliminaries.  Harmonic analysis of scalar functions on
the two-sphere \sphere\ and on the rotation group \sothree\ is
reviewed first, before discussing the analysis of spin functions on
\sphere.  By making all definitions explicit we hope to avoid any
confusion over the conventions adopted.  The reader familiar with
harmonic analysis on \sphere\ and \sothree\ may choose to skip this
section and refer back to it as required.  However, in the final part
of \sectn{\ref{sec:prelim_wigner}} we present some important symmetry
relations of Wigner functions that are required in the derivation
of our algorithms.

\subsection{Scalar spherical harmonics}

The spherical harmonic functions $\shfarg{\el}{\m}{\sa}$, with natural
\mbox{$\el\in\naturals$} and integer $\m\in\integers$ satisfying $|\m|\leq\el$, form an
orthogonal basis in the space of $\ltwo(\sphere,\dmu{\sa})$ scalar
functions on the two-sphere \sphere, where $\sa \equiv (\sas) \in
\sphere$ are the spherical coordinates with co-latitude $\saa\in
[0,\pi]$ and longitude $\sab\in[0,2\pi)$ and $\dmu{\sa}=\sin\saa \dx
\saa \dx \sab$ is the usual rotation invariant measure on the sphere.
The spherical harmonics are defined by
\begin{equation}
\label{eqn:shf}
\shfarg{\el}{\m}{\sas} = 
\sqrt{\frac{2\el+1}{4\pi} \elmfact} \:
\aleg{\el}{\m}{\cos\saa} \:
\exp{\img \m \sab} 
\spcend ,
\end{equation}
where 
\begin{equation}
\aleg{\el}{\m}{x} 
= 
(-1)^\m \: 
\frac{1}{2^\el \el!}  \:
(1-x^2)^{\m/2} \:
\frac{\dx^\m}{\dx x^\m} 
\frac{\dx^\el}{\dx x^\el} \:
(x^2-1)^\el
\end{equation}
are the associated Legendre functions.  We adopt the Condon-Shortley
phase convention, with the $(-1)^\m$ phase factor included in the
definition of the associated Legendre functions, to ensure that the
conjugate symmetry relation $\shfargc{\el}{\m}{\sa} = (-1)^\m
\shfargsp{\el}{-\m}{\sa}$ holds, where the superscript ${}^\cconj$
denotes complex conjugation.  The orthogonality and completeness
relations for the spherical harmonics defined in this manner read
\begin{equation}
\label{eqn:shortho}
\int_{\sphere} 
\dmu{\sa} \:
\shfarg{\el}{\m}{\sa} \:
\shfargc{\el\p}{\m\p}{\sa} 
 = 
\kron{\el}{\el\p}
\kron{\m}{\m\p}
\end{equation}
and
\begin{equation}
\sumlm
\shfarg{\el}{\m}{\sa} \:
\shfargc{\el}{\m}{\sa\p} 
=
\delta(\sa - \sa\p)
\end{equation}
respectively, where $\kron{i}{j}$ is the Kronecker delta, $\delta(\sa
- \sa\p) = \delta(\sab - \sab\p) \: \delta(\cos\saa - \cos\saa\p)$ and
$\delta(x)$ is the Dirac delta function.  Since the spherical harmonic
functions form a complete, orthogonal basis on the sphere, any square
integrable function on the sphere $f\in \ltwo(\sphere,\dmu{\sa})$ may
be represented by the spherical harmonic expansion
\begin{equation}
\label{eqn:sht_inv}
f(\sa) = 
\sum_{\el=0}^\infty
\sum_{\m=-\el}^\el
\shc{f}{\el}{\m} \:
\shfarg{\el}{\m}{\sa}
\spcend ,
\end{equation}
where the spherical harmonic coefficients are given by the usual
projection on to the spherical harmonic basis functions
\begin{equation}
\label{eqn:sht_fwd}
\shc{f}{\el}{\m} =
\int_{\sphere} \dmu{\sa} \:
f(\sa) \:
\shfargc{\el}{\m}{\sa}
\spcend .
\end{equation}
The conjugate symmetry relation of the spherical harmonic coefficients
of a real function (\ie\ $\f^\cconj=\f$) is given by \mbox{$\shcsp{f}{\el}{-\m} = (-1)^\m
\shcc{f}{\el}{\m}$}, which follows directly from the conjugate symmetry
of the spherical harmonic functions.

\subsection{Wigner functions}
\label{sec:prelim_wigner}

We describe here representations of the three-dimensional rotation
group \sothree.  Any rotation $\eul\in\sothree$ is uniquely defined by
the Euler angles $\eul=(\euls)$, where $\eula\in[0,2\pi)$,
$\eulb\in[0,\pi]$ and $\eulc\in[0,2\pi)$.  We adopt the $zyz$ Euler
convention corresponding to the right-handed rotation of a physical body in a fixed
coordinate system about the $z$, $y$ and $z$ axes by \eulc, \eulb\ and
\eula\ respectively (since the axes remain fixed, two rotations about
the $z$ axis are performed).  The Wigner \mbox{$\dmatbig$-functions} \Dlmnp,
with natural $\el\in\naturals$ and integer $\m,\n\in\integers$ satisfying
$|\m|,|\n|\leq\el$, are the matrix elements of the irreducible unitary
representation of the rotation group \sothree.  Moreover, the matrix
elements \Dlmn\ also form an orthogonal basis on
$\ltwo(\sothree,\deul{\eul})$, where $\deul{\eul}=\sin\eulb \dx\eula
\dx\eulb \dx \eulc$ is the invariant measure on the rotation group.
The orthogonality and completeness relations for the Wigner
\dmatbig-functions read
\begin{equation}
  \label{eqn:wigner_ortho}
  \int_\sothree 
  \deul{\eul} \:
  \Dlmnp \:
  \dmatbig_{\m\p\n\p}^{\el\p\cconj}(\eul)
  =
  \frac{8\pi^2}{2\el+1} \:
  \kron{\el}{\el\p} \kron{\m}{\m\p} \kron{\n}{\n\p}
\end{equation}
and
\begin{equation}
  \suml
  \frac{2\el+1}{8\pi^2}
  \summ
  \sumn
  \Dlmnp \:
  \Dlmnc(\eul\p)
  =\delta(\eul-\eul\p)
\end{equation}
respectively, where $\delta(\eul-\eul\p) = \delta(\eula - \eula\p) \,
\delta(\cos\eulb - \cos\eulb\p) \, \delta(\eulc - \eulc\p)$.  Since
the Wigner \dmatbig-functions form a complete, orthogonal basis on the
rotation group \sothree, any square integrable function on the
rotation group \mbox{$F\in \ltwo(\sothree,\deul{\eul})$} may be represented
by the Wigner \dmatbig-function expansion
\begin{equation}
  F(\eul) = 
  \suml
  \frac{2\el+1}{8\pi^2}
  \summ
  \sumn \: F^\el_{\m\n} \: \Dlmnp 
  \spcend ,
\end{equation}
where the Wigner harmonic coefficients are again given in the usual
manner by
\begin{equation}
  F^\el_{\m\n} = \int_\sothree  \deul{\eul} \: F(\eul) \: \Dlmnpc 
  \spcend .
\end{equation} 


The Wigner \dmatbig-functions may be decomposed in terms of the
reduced Wigner \dmatsmall-functions by
\begin{equation}
  \label{eqn:wigner_decomp}
  \Dlmnabg
  = \exp{-\img \m\eula} \:
  \dlmnb \:
  \exp{-\img \n\eulc}
  \spcend .
\end{equation}
The real \dmatsmall-functions may be defined by
\ifjnr
\begin{align}
\label{eqn:wignerd_b}
  \dlmnb =&
  \sqrt{\frac{(\el+\n)! (\el-\n)!}{(\el+\m)! (\el-\m)!}}
  \left( \sin\frac{\eulb}{2} \right)^{\n-\m} \nonumber \\ & \times
  \left( \cos\frac{\eulb}{2} \right)^{\n+\m} 
  \jacobi{\el-\n}{\n-\m}{\n+\m}{\cos\eulb}
  \spcend ,
\end{align}
\else
\begin{equation}
\label{eqn:wignerd_b}
  \dlmnb =&
  \sqrt{\frac{(\el+\n)! (\el-\n)!}{(\el+\m)! (\el-\m)!}}
  \left( \sin\frac{\eulb}{2} \right)^{\n-\m} \nonumber \\ & \times
  \left( \cos\frac{\eulb}{2} \right)^{\n+\m} 
  \jacobi{\el-\n}{\n-\m}{\n+\m}{\cos\eulb}
  \spcend ,
\end{equation}
\fi
where $\jacobi{\el}{a}{b}{\cdot}$ are the Jacobi polynomials.
%
For alternative definitions of the \dmatsmall-functions see the
selection compiled in \cite{mcewen:thesis} and references therein.
Recursion formulae and symmetry relations may be used to compute
rapidly the Wigner functions in the basis of either complex
\cite{risbo:1996,kostelec:2008} or real
\cite{ivanic:1996,blanco:1997} spherical harmonics.  In the
implementations described in this work, we employ the recursion
formulae described in \cite{risbo:1996}.

The \dmatsmall-functions satisfy a number of symmetry relations.  In
this work we make use of the symmetry relations
\begin{equation}
  \label{eqn:wignerd_sym1} 
  \dlmnb = (-1)^{\m-\n} \:  \dmatsmall_{-\m,-\n}^\el(\eulb) \spcend  
  \spcend ,
\end{equation}
\begin{equation}
  \label{eqn:wignerd_sym2} 
  \dlmn(\pi-\eulb) = (-1)^{\el-\n} \: \dmatsmall_{-\m,\n}^\el(\eulb)
\end{equation}
and
\begin{equation}
  \label{eqn:wignerd_sym3} 
  \dlmn(-\eulb) = (-1)^{\m-\n} \: \dlmn(\eulb)
  \spcend ,
\end{equation}
which may be found in \cite{varshalovich:1989}.  In the derivation of
our fast spin spherical harmonic transform we often consider
\dmatsmall-functions of the form \dlmnhalfpi.  For this special case, 
we may infer
\begin{equation}
  \label{eqn:wignerd_sym6} 
  \dlmmnhalfpi = (-1)^{\el+\m} \: \dlmmnnhalfpi
\end{equation}
from \eqn{\ref{eqn:wignerd_sym1}} and \eqn{\ref{eqn:wignerd_sym2}},
which implies that \dlmnhalfpi\ is zero for $\n=0$ and odd $\el+\m$.  We make
continual use of these symmetry relations later in the derivation of
our fast algorithms.

Finally, we note an important Wigner \dmatsmall-function
decomposition.  The \dmatsmall-function for an arbitrary argument may
be decomposed into sums of \mbox{\dmatsmall-functions} for a constant
argument of $\pi/2$ \cite{nikiforov:1991,risbo:1996}:
\begin{equation}
  \label{eqn:wigner_sum_reln}
  \dlmnb = \img^{\n-\m} \sum_{\m\p=-\el}^\el
  \dmatsmall_{\m\p\m}^\el(\pi/2) \:
  \dmatsmall_{\m\p\n}^\el(\pi/2) \:
  \exp{\img \m\p \eulb}
  \spcend ,
\end{equation}
which follows from a factoring of rotations \cite{risbo:1996}.
The Fourier series representation of \dlmnb\ specified by
\eqn{\ref{eqn:wigner_sum_reln}} allows one to write the spherical
harmonic expansion of a function \f\ on \sphere\ in terms of a Fourier
series expansion of \f\ on the two-torus \torus, with \f\ appropriately
extended to this domain (as discussed in more detail in section
\sectn{\ref{sec:algorithms_extension}}).  Consequently,
\eqn{\ref{eqn:wigner_sum_reln}} is fundamental to the derivation of
our fast algorithms.

\subsection{Spin spherical harmonics}

Spin functions on the sphere $\fs\in \ltwo(\sphere, \dmu{\sa})$, with
integer spin $\spin\in\integers$, are defined by their behaviour under
local rotations.  By definition, a spin function transforms as
\begin{equation}
\label{eqn:spin_rot}
\fs^\prime(\sa) = \exp{-\img \spin \chi} \: \fs(\sa)
\end{equation}
under a local rotation by $\chi$, where the prime denotes the rotated
function.  It is important to note that the rotation considered here
is \emph{not} a global rotation on the sphere, such as that
represented by an element of the rotation group \sothree, but rather a
rotation by $\chi$ in the tangent plane at $\sa$.  The sign convention
that we adopt here for the argument of the complex exponential in
\eqn{\ref{eqn:spin_rot}} differs to the original definition
\cite{newman:1966} but is identical to the convention used
recently in the context of the polarisation of the \cmb\
\cite{zaldarriaga:1997}. 

The spin spherical harmonics $\sshfarg{\el}{\m}{\sas}{\spin}$ form an
orthogonal basis for $\ltwo(\sphere, \dmu{\sa})$ spin \spin\ functions
on the sphere for $|\spin|\leq\el$.  Spin spherical harmonics were
first developed by Newman \& Penrose \cite{newman:1966} and were soon
realised to be closely related to the Wigner \dmatbig-functions by
Goldberg \cite{goldberg:1967}.  Spin functions may be defined
equivalently from the expansion in Wigner \dmatbig-functions
$\Dlmn(\euls)$ of a function in $\ltwo(\sothree, \deul{\eul})$
evaluated at $\eulc=0$, for fixed $\n$.
Consequently, the functions 
$\dmatbig_{\m\spin}^{\el}(\sab,\saa  ,0)$
or
$\dmatbig_{\m,-\spin}^{\el\:\cconj}(\sab,\saa  ,0)$, for fixed \spin,
naturally define an orthogonal basis for the expansion of spin \spin\ functions
on the sphere.  The appropriately normalised spin spherical harmonics
may therefore be given by
\begin{equation}
\sshfarg{\el}{\m}{\sas}{\spin} = (-1)^\spin 
\sqrt{\frac{2\el+1}{4\pi} } \:
\dmatbig_{\m,-\spin}^{\el\:\cconj}(\sab,\saa  ,0)
\spcend .
\end{equation}
Noting the decomposition given by \eqn{\ref{eqn:wigner_decomp}}, the
spin spherical harmonics may also be defined in terms of the reduced Wigner
\dmatsmall-functions:
\begin{equation}
\label{eqn:ssh_wigner}
\sshfarg{\el}{\m}{\sas}{\spin} = (-1)^\spin 
\sqrt{\frac{2\el+1}{4\pi} } \:
\dmatsmall_{\m,-\spin}^{\el}(\saa) \:
\exp{\img \m \sab}
\spcend .
\end{equation}
For completeness, we also state the explicit expression derived by
\cite{goldberg:1967} for the spin spherical harmonics:
\ifjnr
\begin{align}
\label{eqn:ssh_goldberg}
&\sshfarg{\el}{\m}{\sas}{\spin} = 
\sqrt{
\frac{2\el+1}{4\pi}
\frac{(\el+\m)!(\el-\m)!}{(\el+\spin)!(\el-\spin)!}
} 
\nonumber \\
&\quad\times
\sin^{2\el}(\saa/2) \:
\exp{\img \m \sab} 
\nonumber \\
&\quad\times
\sum_{r}
\Biggl [ 
\binom{\el-\spin}{r}
\binom{\el+\spin}{\spin-\m+r}
\nonumber \\
& \quad \quad \quad \quad
\times
(-1)^{\el-\spin-r}
\cot^{\spin-\m+2r}(\saa/2)
\Biggr ] 
\spcend ,
\end{align}
\else
\begin{align}
\label{eqn:ssh_goldberg}
\sshfarg{\el}{\m}{\sas}{\spin} = &
\sqrt{
\frac{2\el+1}{4\pi}
\frac{(\el+\m)!(\el-\m)!}{(\el+\spin)!(\el-\spin)!}
} \:
\sin^{2\el}(\saa/2) \:
\exp{\img \m \sab} \nonumber \\
& \times
\sum_{r}
\binom{\el-\spin}{r}
\binom{\el+\spin}{\spin-\m+r}
(-1)^{\el-\spin-r}
\cot^{\spin-\m+2r}(\saa/2)
\spcend ,
\end{align}
\fi
where the sum is performed over all values of $r$ such that the
arguments of the factorials are non-negative.  The scalar spherical
harmonics are identified with the spin spherical harmonics for spin
$\spin=0$, through the relation
\begin{equation}
\dmatsmall_{\m0}^{\el}(\saa) =
\sqrt{\elmfact} \:
\aleg{\el}{\m}{\cos \saa} 
\spcend ,
\end{equation}
\ie\ 
$\sshfarg{\el}{\m}{\sa}{0} = \shfarg{\el}{\m}{\sa}$.
The conjugate symmetry relation given for the scalar spherical
harmonics generalises to the spin spherical harmonics by
$\sshfargc{\el}{\m}{\sa}{\spin} = (-1)^{\spin+\m}
\sshfargsp{\el}{-\m}{\sa}{-\spin}$.

The orthogonality and completeness of the spin spherical
harmonics follows from the orthogonality and completeness of the Wigner
\dmatbig-functions and read
\begin{equation}
\int_{\sphere} 
\dmu{\sa} \:
\sshfarg{\el}{\m}{\sa}{\spin} \:
\sshfargc{\el\p}{\m\p}{\sa}{\spin}
 = 
\kron{\el}{\el\p}
\kron{\m}{\m\p}
\end{equation}
and
\begin{equation}
\sumlm
\sshfarg{\el}{\m}{\sa}{\spin} \:
\sshfargc{\el}{\m}{\sa\p}{\spin}
=
\delta(\sa - \sa\p)
\end{equation}
respectively.
The spin spherical harmonics form a complete, orthogonal basis
for spin functions on the sphere, hence any square integral
spin \spin\ function on the sphere ${}_\spin f\in
\ltwo(\sphere,\dmu{\sa})$ may be represented by the spin \spin\ spherical
harmonic expansion
\begin{equation}
\label{eqn:ssht_inv}
\fs(\sa) = 
\sum_{\el=0}^\infty
\sum_{\m=-\el}^\el
\shc{\fs}{\el}{\m} \:
\sshfarg{\el}{\m}{\sa}{\spin}
\spcend ,
\end{equation}
where the spin spherical harmonic coefficients are given in the usual
manner by 
\begin{equation}
\label{eqn:ssht_fwd}
\shc{\fs}{\el}{\m} =
\int_{\sphere} \dmu{\sa} \,
\fs(\sa) \:
\sshfargc{\el}{\m}{\sa}{\spin}
\spcend .
\end{equation}
The conjugate symmetry relation of the spin spherical harmonic
coefficients is given by ${}_{-\spin}\shcsp{\f}{\el}{-\m} =
(-1)^{\spin+\m} \: \shcc{\fs}{\el}{\m}$ for a function satisfying
\mbox{$\fs^\cconj=\fsm$} (which, for a spin $\spin=0$ function, reduces to
the usual reality condition $\f^\cconj=\f$ for scalar functions) and follows directly from the conjugate
symmetry of the spin spherical harmonics.

Spin raising and lowering operators, $\spinup$ and $\spindown$
respectively, exist so that spin $\spin\pm1$ functions may be defined
from spin \spin\ functions \cite{newman:1966,goldberg:1967}.  When applied to a spin \spin\ function
the resulting function transforms as
$(\spinup \fs)^\prime(\sa) = \exp{-\img (\spin+1) \chi} (\spinup \fs)(\sa)$
and
$(\spindown \fs)^\prime(\sa) = \exp{-\img (\spin-1) \chi} (\spindown \fs)(\sa)$,
where the primed function again corresponds to a rotation by $\chi$ in the
tangent plane at $\sa$.  The spin operators are given explicitly by
\ifjnr
\begin{equation}
\spinup \fs =
-\sin^\spin \saa 
\biggl(\frac{\partial}{\partial\saa} 
+ \frac{\img}{\sin\saa}\frac{\partial}{\partial\sab} \biggr)
\sin^{-\spin} \saa \:
\fs
\end{equation}
\else
\begin{equation}
(\spinup \fs) (\sas) = \biggl[
-\sin^\spin \saa 
\biggl(\frac{\partial}{\partial\saa} 
+ \frac{\img}{\sin\saa}\frac{\partial}{\partial\sab} \biggr)
\sin^{-\spin} \saa \:
\fs \biggr] (\sas)
\end{equation}
\fi
and
\ifjnr
\begin{equation}
\spindown \fs =
-\sin^{-\spin} \saa 
\biggl(\frac{\partial}{\partial\saa} 
- \frac{\img}{\sin\saa}\frac{\partial}{\partial\sab} \biggr)
\sin^{\spin} \saa \:
\fs
\spcend .
\end{equation}
\else
\begin{equation}
(\spindown \fs) (\sas) = \biggl[
-\sin^{-\spin} \saa 
\biggl(\frac{\partial}{\partial\saa} 
- \frac{\img}{\sin\saa}\frac{\partial}{\partial\sab} \biggr)
\sin^{\spin} \saa \:
\fs \biggr] (\sas)
\spcend .
\end{equation}
\fi

\section{Fast algorithms}
\label{sec:algorithms}

The harmonic formulation of our fast spherical harmonic transform
algorithm for spin functions on the sphere is derived here and
contains the essence of the fast algorithm.  The harmonic formulation
involves recasting the spherical harmonic transform on the sphere as a
Fourier series expansion on the torus.  A number of subtleties then
arise due to the method used to extend a function defined on the
sphere to the torus and associated symmetry properties.  After
discussing these subtleties we then present the fast forward an
inverse spin transforms in detail, including a number of practical
issues pertaining to an implementation.

\subsection{Harmonic formulation}
\label{sec:algorithms_harmonic}

Although we initially wish to compute the spherical harmonic
coefficients \fslm\ of a spin \spin\ function \mbox{$\fs\in
\ltwo(\sphere,\dmu{\sa})$}, \ie\ we wish to evaluate the forward
transform specified by \eqn{\ref{eqn:ssht_fwd}}, we will instead
consider the inverse transform specified by \eqn{\ref{eqn:ssht_inv}}.
Using the inverse transform to relate the function \fs\ to its
spherical harmonic coefficients \fslm\ has the advantage that it
doesn't require any quadrature rule to approximate an integral over
the sphere and, consequently, it is exact, up to numerical precision.
It will then be possible to construct a fast and exact direct transform by
inverting the harmonic representation of the fast inverse transform.

Starting with the representation of the spin spherical harmonics in
terms of the reduced Wigner \dmatsmall-functions given by
\eqn{\ref{eqn:ssh_wigner}}, and then substituting the
\dmatsmall-function decomposition given by
\eqn{\ref{eqn:wigner_sum_reln}}, the spin spherical harmonics can be
written in terms of a sum of weighted complex exponentials:
\ifjnr
\begin{align}
\label{eqn:shf_exp}
&\sshfarg{\el}{\m}{\sas}{\spin} =
(-1)^\spin \:
\img^{-(\m+\spin)} \:
\sqrt{\frac{2\el+1}{4\pi} } \:
\nonumber \\
&\quad\times
\sum_{\m\p=-\el}^\el
  \dmatsmall_{\m\p\m}^\el(\pi/2) \:
  \dmatsmall_{\m\p,-\spin}^\el(\pi/2) \:
\exp{\img (\m\sab + \m\p\saa)}
\spcend .
\end{align}
\else
\begin{equation}
\label{eqn:shf_exp}
\sshfarg{\el}{\m}{\sas}{\spin} =
(-1)^\spin \:
\img^{-(\m+\spin)} \:
\sqrt{\frac{2\el+1}{4\pi} } \:
\sum_{\m\p=-\el}^\el
  \dmatsmall_{\m\p\m}^\el(\pi/2) \:
  \dmatsmall_{\m\p,-\spin}^\el(\pi/2) \:
\exp{\img (\m\sab + \m\p\saa)}
\spcend .
\end{equation}
\fi
Substituting this expression for the spin spherical harmonics into the
inverse spin spherical harmonic transform defined by
\eqn{\ref{eqn:ssht_inv}}, we obtain
\ifjnr
\begin{align}
\label{eqn:sht_inv_fast1}
\fs(\sas) = 
\sum_{\el=0}^\elmax
\sum_{\m=-\el}^\el
\fslm \:
(-1)^\spin \:
\img^{-(\m+\spin)} \:
\sqrt{\frac{2\el+1}{4\pi} } \:
\nonumber \\
\times
\sum_{\m\p=-\el}^\el
  \dmatsmall_{\m\p\m}^\el(\pi/2) \:
  \dmatsmall_{\m\p,-\spin}^\el(\pi/2) \:
\exp{\img (\m\sab + \m\p\saa)}
\spcend ,
\end{align}
\else
\begin{equation}
\label{eqn:sht_inv_fast1}
\fs(\sas) = 
\sum_{\el=0}^\elmax
\sum_{\m=-\el}^\el
\fslm \:
(-1)^\spin \:
\img^{-(\m+\spin)} \:
\sqrt{\frac{2\el+1}{4\pi} } \:
\sum_{\m\p=-\el}^\el
  \dmatsmall_{\m\p\m}^\el(\pi/2) \:
  \dmatsmall_{\m\p,-\spin}^\el(\pi/2) \:
\exp{\img (\m\sab + \m\p\saa)}
\spcend ,
\end{equation}
\fi
where we assume that \fs\ is band-limited at \elmax, that is \mbox{$\fslm=0$},
$\forall \el > \elmax$, so that the outer summation may be truncated
to \elmax. 
%
%
%
After interchanging the orders of summation,
\eqn{\ref{eqn:sht_inv_fast1}} may be written as
\begin{equation}
\label{eqn:sht_inv_fast}
\fs(\sas) = 
\sum_{\m=-\elmax}^\elmax
\sum_{\m\p=-\elmax}^\elmax
\Fsfourier \:
\exp{\img (\m\sab + \m\p\saa)}
\spcend ,
\end{equation}
where 
\ifjnr
\begin{align}
\label{eqn:fourier_coeff}
&\Fsfourier = 
(-1)^\spin \:
\img^{-(\m+\spin)} 
\nonumber \\
&\quad\times
\sum_{\el=0}^\elmax
\sqrt{\frac{2\el+1}{4\pi} } \:
  \dmatsmall_{\m\p\m}^\el(\pi/2) \:
  \dmatsmall_{\m\p,-\spin}^\el(\pi/2) \:
\fslm
\spcend .
\end{align}
\else
\begin{equation}
\label{eqn:fourier_coeff}
\Fsfourier = 
(-1)^\spin \:
\img^{-(\m+\spin)} 
\sum_{\el=0}^\elmax
\sqrt{\frac{2\el+1}{4\pi} } \:
  \dmatsmall_{\m\p\m}^\el(\pi/2) \:
  \dmatsmall_{\m\p,-\spin}^\el(\pi/2) \:
\fslm
\spcend .
\end{equation}
\fi 
We adopt the convention that all terms indexed by \el\ and an
azimuthal index such as \m\ are zero for $|\m|>\el$.  This convention
may also be represented explicitly by replacing the lower limit of the
\el\ summation of $0$ in \eqn{\ref{eqn:fourier_coeff}} with
$\max(|\m|, |\m\p|)$.

Notice that \eqn{\ref{eqn:sht_inv_fast}} is very similar to the
Fourier series representation of \fs.  However, the Fourier series
expansion is only defined for functions that are periodic, \ie\ for
functions defined on the torus.  A function \fs\ on the sphere is
periodic in \sab\ but not in \saa.  In order to apply
\eqn{\ref{eqn:sht_inv_fast}} directly it is necessary to extend \fs,
initially defined on the two-sphere \sphere, to be defined on the
two-torus \torus.  Once \fs\ is extended in this manner an inverse
\fft\ may be used to evaluate \Fsfourier\ rapidly from a sampled
version of \fs.  The harmonic coefficients \fslm\ may then be computed
from \Fsfourier\ by inverting \eqn{\ref{eqn:fourier_coeff}} to give a
fast forward spin spherical harmonic transform.  Similarly, \fs\ may
be reconstructed by computing \Fsfourier\ from the harmonic
coefficients \fslm\ through \eqn{\ref{eqn:fourier_coeff}}, followed by
the use of a forward \fft\ to recover a sampled version of \fs\ from
\Fsfourier.  This gives a fast inverse spin spherical harmonic
transform. 
By recasting the spherical harmonic expansion on the sphere as a
Fourier expansion on the torus we may appeal to the exact quadrature
of the Fourier integral on uniformly sampled grids, \ie\ the Shannon
sampling theorem \cite{shannon:1949}.  Consequently, our fast transforms are
theoretically exact (to numerical precision and stability) for a
sampling of the sphere compatible with a uniform grid on the torus (a
corresponding sampling of the sphere is made explicit in
\sectn{\ref{sec:algorithms_fourier_coeff}}).
We address the periodic extension of \fs\ from \sphere\ to
\torus\ next.

\subsection{Extending the sphere to the torus}
\label{sec:algorithms_extension}

When periodically extending \fs\ from \sphere\ to \torus\ it is important
to ensure that all relations used previously are still satisfied on
the new domain.  Careful attention must therefore be paid to the
periodic extension of \fs, as a naive extension may introduce
inconsistency.

We extend \fs\ to $\fst \in \ltwo(\torus, \dx^2\vect{x})$ by allowing
\saa\ to range over the domain $[-\pi, \pi)$ and assuming \fst\ is
periodic with period $2\pi$ in both \sab\ and \saa\ (note that $\dx^2
\vect{x}=\dx\saa \dx\sab$ is the invariant measure on the
$(\sas)$-plane).  Equivalently, we may represent $\theta \in [0,
2\pi)$ and both domains are used interchangeably depending on which is
most convenient.  Representing \fs\ over this extended domain is
redundant, covering the sphere exactly twice.  Nevertheless, such a
representation facilitates our fast algorithm, which more than
compensates for this redundancy.  Such an approach is not uncommon, as
\cite{driscoll:1994,healy:2003} also oversample a function on the
sphere in the \saa\ direction in order to develop alternative fast
scalar spherical harmonic transforms.

The method used to extend \fs\ to the new $2\pi$ \saa\ domain remains
to be chosen.  Two natural candidates are to extend \fs\ in \saa\ in
either an even or odd manner about zero.  We obtain two candidate
periodic extensions of \fs\ on \torus: \fste\ and \fsto, where the
superscripts denote the even and odd extensions respectively, \ie\
$\fste(-\sas) = \fste(\sas)$ and $\fsto(-\sas) = -\fsto(\sas)$.  The
appropriate periodic extension to use depends on the symmetry properties
of the basis functions used to represent the inverse spherical
harmonic transform.  
In \fig{\ref{fig:sphere_to_torus}} we illustrate these periodic
extensions of a function on the sphere to give a function defined on
the torus.  We use an Earth topography map to define our initial
function on the sphere in this example. 

Let us recall the inverse spin spherical harmonic transform given by
\eqn{\ref{eqn:ssht_inv}}.
To ensure consistency, we should extend \fs\ to \torus\ so that it
satisfies the same symmetry properties as $\sshf{\el}{\m}{\spin}$
would when extended to this domain. For the spin $\spin=0$ case, one
might naively expect that $\shf{\el}{\m}$ would exhibit even symmetry
in \saa\ about zero when extended to $\saa\in[-\pi,\pi)$ based on the
definition of $\shf{\el}{\m}$ given by \eqn{\ref{eqn:shf}}.  However,
in deriving the alternative harmonic formulation of the inverse
spherical harmonic transform in \sectn{\ref{sec:algorithms_harmonic}}
we used the representation of $\sshf{\el}{\m}{\spin}$ specified by
\eqn{\ref{eqn:ssh_wigner}}, or equivalently
\eqn{\ref{eqn:ssh_goldberg}}.  Although all of these definitions are
obviously identical on the sphere \sphere, they are \emph{not}
necessarily identical on the torus \torus.  Since we have used
\eqn{\ref{eqn:ssh_wigner}} in our harmonic formulation we must be sure
to satisfy the symmetry properties of this relation on \torus.  The
symmetry of \eqn{\ref{eqn:ssh_wigner}} in \saa\ is identical to the
symmetry of $\dlms(\saa)$. Recall from \eqn{\ref{eqn:wignerd_sym3}}
that
\begin{equation}
\label{eqn:wignerd_sym3a}
\dlms(-\saa) = (-1)^{\m+\spin} \:
\dlms(\saa)
\spcend ,
\end{equation}
\ie\ $\dlms(\saa)$ is even for even $\m+\spin$ and odd for odd
$\m+\spin$. This same symmetry property is also apparent from
\eqn{\ref{eqn:ssh_goldberg}}.\footnote{Here the symmetry of
  $\sshf{\el}{\m}{\spin}$ is dictated by the symmetry of the
  $\cot(\saa/2)$ term.  Since $\cot$ is an odd function the overall
  symmetry depends on the power that $\cot(\saa/2)$ is raised to.  All
  $\cot$ terms in the sum have either an even or odd power since the
  power increases by $2r$ for each term.  The overall symmetry
  therefore depends on the parity of $\spin-\m$, or equivalently
  $\spin+\m$.  Thus $\sshf{\el}{\m}{\spin}$ is even for even
  $\spin+\m$ and odd for odd $\spin+\m$.}  
This symmetry could naively appear to be problematic.  There is no way
we can extend \fs\ to \torus\ to satisfy this relation.  However, we
may solve this problem by considering both \fte\ and \fto.  For even
$\m+\spin$, \eqn{\ref{eqn:wignerd_sym3a}} will be satisfied for \fste,
whereas for odd $\m+\spin$ it will be satisfied for \fsto.

Our fast spherical harmonic transform then proceeds as follows.  We
perform two inverse \fft s to compute rapidly \Fsfouriere\ and
\Fsfouriero\ for both \fste\ and \fsto\ respectively.  We then invert
\eqn{\ref{eqn:fourier_coeff}} to compute the harmonic coefficients
\fstelm\ and \fstolm\ corresponding to \fste\ and \fsto\ respectively.
The harmonic coefficients \fslm\ of \fs\ will be identical to \fstelm\
for $\m+\spin$ even only, and the identical to \fstolm\ for $\m+\spin$
odd only, \ie
\begin{equation}
\fslm=
\begin{cases}
\fstelm & \m+\spin \mbox{ even}\\
\fstolm & \m+\spin \mbox{ odd}\\
\end{cases} 
\spcend .
\end{equation}

\newlength{\sphereheight}
\setlength{\sphereheight}{35mm}

\newlength{\torusheight}
\setlength{\torusheight}{35mm}

\begin{figure*}[t]
\centering
\centering
\mbox{
\subfigure[$\fs\in\ltwo(\sphere,\dmu{\sa})$]{\includegraphics[height=\sphereheight]{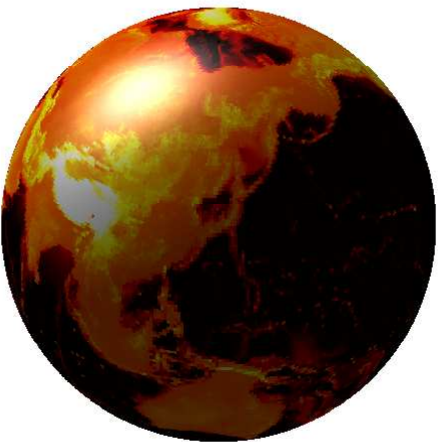}}
\quad\quad \quad\quad
\subfigure[$\fste\in \ltwo(\torus, \dx^2\vect{x})$]{\includegraphics[height=\torusheight]{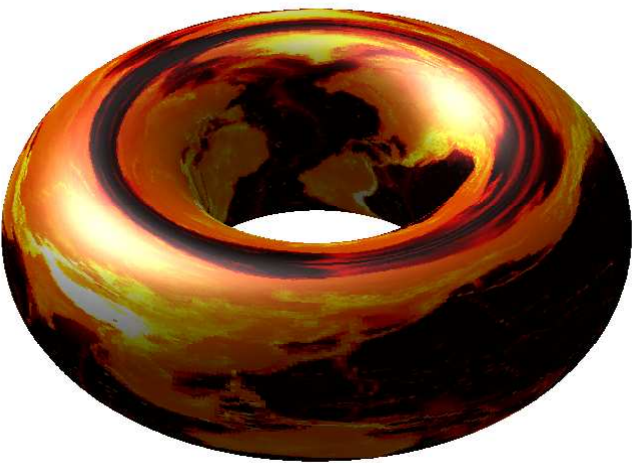}}
\quad\quad \quad\quad
\subfigure[$\fsto\in \ltwo(\torus, \dx^2\vect{x})$]{\includegraphics[height=\torusheight]{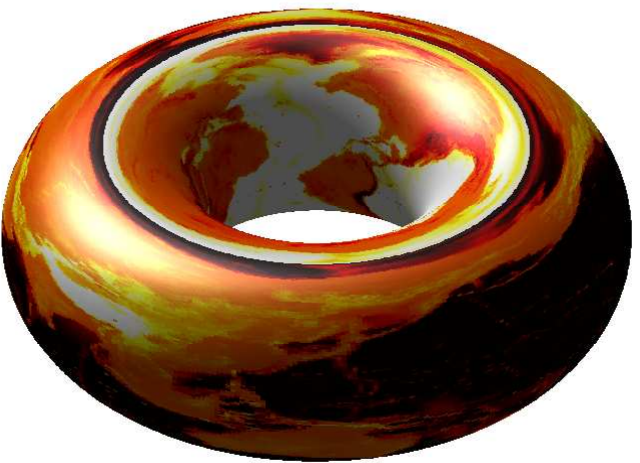}}
}
\caption{Periodic extension to the torus of Earth topography data
  defined originally on the sphere (mountains are shown in white and
  lakes in dark red/grey, with the colour of intermediate values interpolated
  between these extremes).  The left panel shows the Earth
  data defined on the sphere, while the centre panel shows the even
  periodic extension of the data to the torus and the right panel shows
  the odd periodic extension.  Both types of periodic extension are used
  in our fast algorithm.}
\label{fig:sphere_to_torus}
\end{figure*}

\subsection{Symmetry relations}

The Fourier coefficients \Fsfourier\ satisfy a number of symmetry
relations which may be exploited to increase the speed of
implementations.  Here we examine the symmetry properties of
\Fsfourier\ using two separate approaches.  Firstly, we determine the
symmetries of \Fsfouriere\ and \Fsfouriero\ based on the Fourier
series representations of \fste\ and \fsto.  Secondly, we examine the
symmetries of \Fsfourier\ computed directly from \fslm\ using
\eqn{\ref{eqn:fourier_coeff}}.  Recall that the symmetries of
\Fsfourier\ will only be satisfied by \Fsfouriere\ for $\m+\spin$ even
and by \Fsfouriero\ for $\m+\spin$ odd.

Although in practice we will compute \Fsfouriere\ and \Fsfouriero\ by
taking \fft s of sampled versions of \fste\ and \fsto\ respectively, we
may nevertheless represent \Fsfouriere\ and \Fsfouriero\ by the usual
projection of the continuous functions on to the complex exponentials,
\ie\
\begin{equation}
\label{eqn:fourier_series}
\Fstfourier = \frac{1}{(2\pi)^2} 
\int_0^{2\pi} \dx \saa 
\int_0^{2\pi} \dx \sab \:
\fst(\sas) \:
\exp{-\img (\m\sab + \m\p\saa) }
\spcend ,
\end{equation}
where, here and subsequently, we drop the superscript denoting the
even or odd function to indicate that the relation holds for both (in
fact, the discrete Fourier transform merely gives an exact quadrature
rule for evaluating \eqn{\ref{eqn:fourier_series}}).  By considering
all combinations of negative \m\ and $\m\p$ we obtain trivially the
following symmetry relations for \Fsfouriere from
\eqn{\ref{eqn:fourier_series}}:
\begin{align}
 {}_{\spin} \Ft_{\m,-\m\p}^{\rm e} &= \Fst_{\m\m\p}^{\rm e} ; \\
 {}_{-\spin} \Ft_{-\m,\m\p}^{\rm e} &= \Fst_{\m\m\p}^{\rm e}{}^{\cconj} \quad \mbox{for $\fs^\cconj=\fsm$} ; \\
 {}_{-\spin} \Ft_{-\m,-\m\p}^{\rm e} &= \Fst_{\m\m\p}^{\rm e}{}^{\cconj}\quad \mbox{for $\fs^\cconj=\fsm$} .
\end{align}
Similarly, we obtain the following symmetry relations for \Fsfouriero:
\begin{align}
 {}_{\spin} \Ft_{\m,-\m\p}^{\rm o} &= (-1)\: \Fst_{\m\m\p}^{\rm o} ; \\
 {}_{-\spin} \Ft_{-\m,\m\p}^{\rm o} &= (-1)\: \Fst_{\m\m\p}^{\rm o}{}^{\cconj} \quad \mbox{for $\fs^\cconj=\fsm$} ; \\
 {}_{-\spin} \Ft_{-\m,-\m\p}^{\rm o} &=  \Fst_{\m\m\p}^{\rm o}{}^{\cconj} \quad \mbox{for $\fs^\cconj=\fsm$} .
\end{align}

Now we derive symmetry relations for \Fsfourier\ computed directly
from \fslm\ through \eqn{\ref{eqn:fourier_coeff}}.  Firstly, we find
\ifjnr
\begin{align}
\label{eqn:Ffourier_sym}
\Fs_{\m,-\m\p} 
&=
(-1)^\spin \:
\img^{-(\m+\spin)} 
\nonumber \\
&\quad\times
\sum_{\el=0}^\elmax
\Biggl [
\sqrt{\frac{2\el+1}{4\pi} } \:
  \dmatsmall_{-\m\p,\m}^\el(\pi/2) \:
\nonumber \\
&\quad\quad\quad\quad\times
  \dmatsmall_{-\m\p,-\spin}^\el(\pi/2) \:
\fslm \Biggr ]
\nonumber \\
 &=
(-1)^\spin \:
\img^{-(\m+\spin)} 
\nonumber \\
&\quad\times
\sum_{\el=0}^\elmax
\Biggl [
\sqrt{\frac{2\el+1}{4\pi} } \:
  (-1)^{\el+\m} \: \dmatsmall_{\m\p\m}^\el(\pi/2) \:
\nonumber \\
&\quad\quad\quad\quad\times
  (-1)^{\el-\spin} \: \dmatsmall_{\m\p,-\spin}^\el(\pi/2) \:
\fslm \Biggr ]
\nonumber \\
&= (-1)^{\m+\spin} \: \Fs_{\m\m\p} 
\spcend ,
\end{align}
\else
\begin{align}
\label{eqn:Ffourier_sym}
\Fs_{\m,-\m\p} 
&=
(-1)^\spin \:
\img^{-(\m+\spin)} 
\sum_{\el=0}^\elmax
\sqrt{\frac{2\el+1}{4\pi} } \:
  \dmatsmall_{-\m\p,\m}^\el(\pi/2) \:
  \dmatsmall_{-\m\p,-\spin}^\el(\pi/2) \:
\fslm \nonumber \\
 &=
(-1)^\spin \:
\img^{-(\m+\spin)} 
\sum_{\el=0}^\elmax
\sqrt{\frac{2\el+1}{4\pi} } \:
  (-1)^{\el+\m} \: \dmatsmall_{\m\p\m}^\el(\pi/2) \:
  (-1)^{\el-\spin} \: \dmatsmall_{\m\p,-\spin}^\el(\pi/2) \:
\fslm \nonumber \\
&= (-1)^{\m+\spin} \: \Fs_{\m\m\p} 
\spcend ,
\end{align}
\fi
where in the second line we have applied the \dmatsmall-function
symmetry relation \eqn{\ref{eqn:wignerd_sym2}}.  For the second
symmetry relation we find
\ifjnr
\begin{align}
{}_{-\spin} \F_{-\m,\m\p} 
&=
(-1)^\spin \:
\img^{\m+\spin} 
\nonumber \\
&\quad\times
\sum_{\el=0}^\elmax
\Biggl [
\sqrt{\frac{2\el+1}{4\pi} } \:
  \dmatsmall_{\m\p,-\m}^\el(\pi/2) \:
\nonumber \\
&\quad\quad\quad\quad\times
  \dmatsmall_{\m\p\spin}^\el(\pi/2) \:
\sshcsp{\f}{\el}{-\m}{-\spin} \Biggr ]
\nonumber \\
&=
(-1)^\spin \:
(\img^\cconj)^{-(\m+\spin)}  
\nonumber \\
&\quad\times
\sum_{\el=0}^\elmax
\Biggl [
\sqrt{\frac{2\el+1}{4\pi} } \:
  \dmatsmall_{\m\p\m}^\el(\pi/2) \:
\nonumber \\
&\quad\quad\quad\quad\times
  \dmatsmall_{\m\p,-\spin}^\el(\pi/2) \:
(-1)^{\m+\spin} \sshcc{\f}{\el}{\m}{\spin} 
\Biggr ]
\nonumber \\
&= (-1)^{\m+\spin} \: \Fs_{\m\m\p}{}^\cconj \quad \mbox{for $\fs^\cconj=\fsm$},
\end{align}
\else
\begin{align}
{}_{-\spin} \F_{-\m,\m\p} 
&=
(-1)^\spin \:
\img^{\m+\spin} 
\sum_{\el=0}^\elmax
\sqrt{\frac{2\el+1}{4\pi} } \:
  \dmatsmall_{\m\p,-\m}^\el(\pi/2) \:
  \dmatsmall_{\m\p\spin}^\el(\pi/2) \:
\sshcsp{\f}{\el}{-\m}{-\spin} \nonumber \\
&=
(-1)^\spin \:
(\img^\cconj)^{-(\m+\spin)}  
\sum_{\el=0}^\elmax
\sqrt{\frac{2\el+1}{4\pi} } \:
  \dmatsmall_{\m\p\m}^\el(\pi/2) \:
  \dmatsmall_{\m\p,-\spin}^\el(\pi/2) \:
(-1)^{\m+\spin} \sshcc{\f}{\el}{\m}{\spin}  \quad \mbox{for $\fs^\cconj=\fsm$} \nonumber \\
&= (-1)^{\m+\spin} \: \Fs_{\m\m\p}{}^\cconj \quad \mbox{for $\fs^\cconj=\fsm$},
\end{align}
\fi
where in the second line we have applied the \dmatsmall-function
symmetry relations \eqn{\ref{eqn:wignerd_sym1}} and
\eqn{\ref{eqn:wignerd_sym2}}, and noted the harmonic conjugate
symmetry relation for $\fs^\cconj=\fsm$.  It is possible to derive the
third symmetry relation for \Fsfourier\ in a similar manner to the
previous two, however it is much easier to use these two previously
derived relations directly, in which case it may be shown trivially
that
\begin{equation}
{}_{-\spin} \F_{-\m,-\m\p} = {}_{\spin} \F_{\m\m\p} {}^\cconj \quad \mbox{for $\fs^\cconj=\fsm$}.
\end{equation}
Notice that, as expected, we find that the symmetry relations of
\Fsfourier\ are only satisfied by \Fsfouriere\ for $\m+\spin$ even and by
\Fsfouriero\ for $\m+\spin$ odd.

For a complex signal we have a single symmetry relation, which may be
exploited to speed up an implementation by a factor of two.  For a signal
satisfying $\fs^\cconj=\fsm$ (\ie\ a real signal for the case
$\spin=0$), we have two symmetry relations, which may be exploited to
speed up an implementation by a factor of four.  These enhancements
are of practical importance but do not change the overal scaling of
our algorithms.

\subsection{Forward and inverse transform}
\label{sec:algorithms_fwdinv}

The harmonic formulation of our fast algorithm was derived in
\sectn{\ref{sec:algorithms_harmonic}}.  Since this derivation a number
of practicalities have been discussed, such as extending \fs\ defined
on \sphere\ to give \fst\ defined on \torus, and the symmetry
properties of the Fourier coefficients of \fst\ on \torus.  We are now
in a position to describe our fast algorithm in detail.  Firstly, we
consider the discretisation of \fst\ on \torus\ and the inversion of
\eqn{\ref{eqn:sht_inv_fast}} using an \fft\ to compute \Fstfourier\
rapidly.  We then consider the inversion of
\eqn{\ref{eqn:fourier_coeff}} to compute the spherical harmonic
coefficients \fslm.  Finally, we discuss how our formulation may also
be used to compute the inverse spherical harmonic transform, \ie\ to
recover \fs\ from the harmonic coefficients \fslm.  \mbox{Although} our fast
forward and inverse spin spherical harmonic transforms appear to
follow directly from \eqn{\ref{eqn:sht_inv_fast}} and
\eqn{\ref{eqn:fourier_coeff}}, a number of subtleties arise due to the
symmetries of the functions involved.

\subsubsection{Computing Fourier coefficients on the torus}
\label{sec:algorithms_fourier_coeff}

We first discretise \fst\ defined on the two-torus \torus.  For now we
are not concerned with the method by which \fs\ is periodically extended
to \torus, \ie\ through either even or odd symmetry, hence we drop the
related superscript.
Previously, we extended \fst\ to \torus\ by extending the \saa\ domain
to $[-\pi,\pi)$.  However, we noted that \fst\ could equivalently be
represented on the domain $\saa \in [0,2\pi)$.  For convenience, we
choose this second domain here since it simplifies the subsequent
discretisation.  We adopt the equi-angular sampled grid with nodes
given at
\begin{equation}
\label{eqn:nodes_theta}
\saa_\saai = 
\frac{\pi(2\saai+1)}{2\elmax+1}, 
\quad \mbox{where } \saai = 0,1,\dotsc,2\elmax
\end{equation}
and
\begin{equation}
\label{eqn:nodes_phi}
\sab_\sabi = 
\frac{\pi(2\sabi+1)}{2\elmax+1}, 
\quad \mbox{where } \sabi = 0,1,\dotsc,2\elmax
\spcend .
\end{equation}
An odd number of sample points are required in both \saa\ and \sab\ so
that a direct association may be made between the sampled \fst\ and
\Fstfourier\ (where both $\m$ and $\m\p$ range from
$-\elmax,\dotsc,0,\dotsc,\elmax$).  Moreover, we make the association
$N=2\elmax+1$, where $N$ is the number of samples in both the \saa\
and \sab\ dimensions, to ensure that Nyquist sampling
\cite{shannon:1949} is satisfied.
The node positions specified by \eqn{\ref{eqn:nodes_theta}} and
\eqn{\ref{eqn:nodes_phi}} eliminate repeated samples at the poles
$\saa=0$ and $\saa=2\pi$, since these points are excluded from the
grid.  However, it is not possible to eliminate repeated samples at
$\saa=\pi$, since we require a discretisation that is symmetric about
$\pi$ but which contains an odd number of sample points.
The sampled version of \fst\ is given by 
\begin{equation}
 \fst[\sais] = \fst(\saa_\saai, \sab_\sabi)
 \spcend ,
\end{equation}
where we use the convention that square brackets are used to index the
discretised function.

Now that we have sampled \fst, we may invert
\eqn{\ref{eqn:sht_inv_fast}} using an \fft.  This process is
straightforward but we nevertheless discuss the various approaches
that may be used to solve this problem.  We start by discretising
\eqn{\ref{eqn:sht_inv_fast}}:
\ifjnr
\begin{align}
\label{eqn:discrete1}
& \fst[\sais] =\nonumber \\
&\sum_{\m=-\elmax}^\elmax
\sum_{\m\p=-\elmax}^\elmax
\Fstfourier \:
\exp{\img \pi [\m(2\sabi+1) + \m\p(2\saai+1)]/(2\elmax+1)}
\spcend .
\end{align}
\else
\begin{equation}
\label{eqn:discrete1}
 \fst[\sais] = 
\sum_{\m=-\elmax}^\elmax
\sum_{\m\p=-\elmax}^\elmax
\Fstfourier \:
\exp{\img \pi [\m(2\sabi+1) + \m\p(2\saai+1)]/(2\elmax+1)}
\spcend .
\end{equation}
\fi
It is necessary to rewrite \eqn{\ref{eqn:discrete1}} in a form that is
suitable for the application of standard \fft\ algorithms, where we
adopt the forward discretise Fourier transform definition
\ifjnr
\begin{align}
\label{eqn:dft}
G_{\m\m\p} &= 
\ffts \{g[\sais]\} 
\nonumber \\
&= 
\sum_{\sabi=0}^{N-1} \sum_{\saai=0}^{N-1} g[\sais] \: \exp{-\img 2 \pi (\m \sabi + \m\p \saai) / N}
\spcend ,
\end{align}
\else
\begin{equation}
\label{eqn:dft}
G_{\m\m\p} =
\ffts \{g[\sais]\} = 
\sum_{\sabi=0}^{N-1} \sum_{\saai=0}^{N-1} g[\sais] \: \exp{-\img 2 \pi (\m \sabi + \m\p \saai) / N}
\spcend ,
\end{equation}
\fi
with inverse
\ifjnr
\begin{align}
\label{eqn:idft}
g[\sais] &=
\iffts \{G_{\m\m\p}\}
\nonumber \\
&= 
\frac{1}{N^2}
\sum_{\m=0}^{N-1} \sum_{\m\p=0}^{N-1} G_{\m\m\p} \: \exp{\img 2 \pi (\m \sabi + \m\p \saai)/N}
\spcend ,
\end{align}
\else
\begin{equation}
\label{eqn:idft}
g[\sais] =
\iffts \{G_{\m\m\p}\} = 
\frac{1}{N^2}
\sum_{\m=0}^{N-1} \sum_{\m\p=0}^{N-1} G_{\m\m\p} \: \exp{\img 2 \pi (\m \sabi + \m\p \saai)/N}
\spcend ,
\end{equation}
\fi
for indices $\sabi,\:\saai,\:\m,\:\m\p$ all ranging over $0,1,\dotsc,N-1$.  In
order to convert \eqn{\ref{eqn:discrete1}} into a form similar to
\eqn{\ref{eqn:idft}} so that an \fft\ may be applied to evaluate it, we
first shift the summation indices to give
\ifjnr
\begin{align}
&\fst[\sais] = 
\exp{-\img 2 \pi \elmax (\sabi + \saai + 1)/(2\elmax+1)}
\nonumber \\
&\sum_{\m=0}^{2\elmax}
\sum_{\m\p=0}^{2\elmax}
\Bigl [
\Fst_{\m-\elmax,\m\p-\elmax} \:
\exp{\img \pi (\m + \m\p)/(2\elmax+1)}
\nonumber \\
& \quad\quad\quad\quad\quad\times \exp{\img 2 \pi (\m\sabi + \m\p\saai)/(2\elmax+1)}
\Bigr ]
\spcend ,
\end{align}
\else
\begin{equation}
 \fst[\sais] = 
\exp{-\img 2 \pi \elmax (\sabi + \saai + 1)/(2\elmax+1)}
\sum_{\m=0}^{2\elmax}
\sum_{\m\p=0}^{2\elmax}
\Fst_{\m-\elmax,\m\p-\elmax} \:
\exp{\img \pi (\m + \m\p)/(2\elmax+1)}
\exp{\img 2 \pi (\m\sabi + \m\p\saai)/(2\elmax+1)}
\spcend ,
\end{equation}
\fi
from which it follows that
\ifjnr
\begin{align}
&\exp{\img 2 \pi \elmax (\sabi + \saai + 1)/(2\elmax+1)} \:
\fst[\sais] 
=
(2\elmax+1)^2
\nonumber \\
&\quad\times 
\iffts 
\Bigl\{ 
  \exp{\img \pi (\m + \m\p)/(2\elmax+1)} \:
  \Fst_{\m-\elmax,\m\p-\elmax} 
\Bigl\}
\spcend .
\end{align}
\else
\begin{equation}
\exp{\img 2 \pi \elmax (\sabi + \saai + 1)/(2\elmax+1)} \:
\fst[\sais] 
=
(2\elmax+1)^2 \: \iffts 
\Bigl\{ 
  \exp{\img \pi (\m + \m\p)/(2\elmax+1)} \:
  \Fst_{\m-\elmax,\m\p-\elmax} 
\Bigl\}
\spcend .
\end{equation}
\fi
This may be easily inverted using the definition of the discrete
Fourier transform, giving
\ifjnr
\begin{align}
\label{eqn:dft_phase}
&\Fst_{\m-\elmax,\m\p-\elmax} \:
=
\frac{\exp{-\img \pi (\m + \m\p)/(2\elmax+1)}}{(2\elmax+1)^2} \:
\nonumber \\
&\quad\quad\quad\times
\ffts
\Bigl\{
  \exp{\img 2 \pi \elmax (\sabi + \saai + 1)/(2\elmax+1)} \:
  \fst[\sais]
\Bigr\}
\spcend .
\end{align}
\else
\begin{equation}
\label{eqn:dft_phase}
\Fst_{\m-\elmax,\m\p-\elmax} \:
=
\frac{\exp{-\img \pi (\m + \m\p)/(2\elmax+1)}}{(2\elmax+1)^2} \:
\ffts
\Bigl\{
  \exp{\img 2 \pi \elmax (\sabi + \saai + 1)/(2\elmax+1)} \:
  \fst[\sais]
\Bigr\}
\spcend ,
\end{equation}
\fi
which may be evaluated through an \fft\ directly, applying the
appropriate phase shifts at each stage to recover
$\Fst_{\m-\elmax,\m\p-\elmax}$.  Alternatively, we may represent the
phase shift applied to $\fst[\sais]$ in the spatial domain as a spatial
shift in the frequency domain.  In this case we recover
\begin{equation}
\label{eqn:dft_spatial}
\Fst_{\m\m\p}
=
\frac{\exp{-\img \pi (\m + \m\p)/(2\elmax+1)}}{(2\elmax+1)^2} \:
\fftshift
\ffts
\Bigl\{
  \fst[\sais]
\Bigr\}
\spcend ,
\end{equation}
where $\m$ and $\m\p$ now range from $-\elmax,\dotsc,0,\dotsc,\elmax$
and the \fft-shift operator $\fftshift$ interchanges quadrants of a
sampled two-dimensional signal by interchanging quadrant I with IV and
also interchanging quadrant II with III.  The final representation
given by \eqn{\ref{eqn:dft_spatial}} is marginally less
computationally demanding than the representation given by
\eqn{\ref{eqn:dft_phase}} since it is not necessary to compute and
apply one of the phase shifts.  Moreover, \eqn{\ref{eqn:dft_spatial}}
involves an \fft\ of a real signal for real spin $\spin=0$ functions, rather
than a complex one, for which \fft s are approximately twice as fast and
require half the memory requirements.  In our implementation we use
the expression given by \eqn{\ref{eqn:dft_spatial}} to compute
\Fstfourier\ rapidly using an \fft.  Finally, we note that
\eqn{\ref{eqn:dft_spatial}} may be inverted easily through
\ifjnr
\begin{align}
\label{eqn:idft_spatial}
&\fst[\sais]
=
(2\elmax+1)^2 
\nonumber \\
&\quad\quad\times
\iffts
\ifftshift
\Bigl\{
  \exp{\img \pi (\m + \m\p)/(2\elmax+1)} \:
  \Fst_{\m\m\p}
\Bigr\}
\spcend ,
\end{align}
\else
\begin{equation}
\label{eqn:idft_spatial}
\fst[\sais]
=
(2\elmax+1)^2 \:
\iffts
\ifftshift
\Bigl\{
  \exp{\img \pi (\m + \m\p)/(2\elmax+1)} \:
  \Fst_{\m\m\p}
\Bigr\}
\spcend ,
\end{equation}
\fi
where \ifftshift\ is the inverse \fft-shift operator.

We have so far ignored the fact that \fst\ always exhibits either even
or odd symmetry.  This property may be exploited by using fast cosine
or sine transforms in place of the \fft, which will improve the speed
of the algorithms but will not alter their complexity.  For now, we
ignore this optimisation but intend to readdress it in future.

\ifjnr
\fi

\ifjnr
\end{multicols}
\fi

\subsubsection{Forward transform}

Now that we have discussed how to compute \Fstfourier\ rapidly from a
sampled version of \fst, we turn our attention to inverting
\eqn{\ref{eqn:fourier_coeff}} in order to recover the harmonic
coefficients \fstlm.  Recall that \eqn{\ref{eqn:fourier_coeff}} is not
satisfied simultaneously by the two periodic extensions of \fs\ on
\torus, \ie\ \fste\ and \fsto, for all $\m+\spin$.  Instead, it is
satisfied by \fste\ for even $\m+\spin$ and by \fsto\ for odd
$\m+\spin$.  To recover the harmonic coefficients \fslm, we solve
\eqn{\ref{eqn:fourier_coeff}} for \fstelm\ for even $\m+\spin$ and
solve for \fstolm\ for odd $\m+\spin$.  

Let us consider fixed $\m=\mz$ but all $\m\p$. Throughout we
also consider fixed \spin.  For this case we may rewrite
\eqn{\ref{eqn:fourier_coeff}} as the matrix equation
%
\begin{equation*}
\begin{pmatrix}
 \Fst_{\mz,-\elmax} \\
  \vdots \\
 \Fst_{\mz,0} \\
  \vdots \\
 \Fst_{\mz,\elmax} \\
\end{pmatrix}
= \img^{-(\mz+\spin)}
\begin{pmatrix}
                                                                                       &&& \sqrt{\frac{2\elmax+1}{4\pi}}\, \dlmnarg{\elmax}{-\elmax}{\mz}\,\dlmnarg{\elmax}{-\elmax}{-\spin} \\
&&\iddots&\vdots\\
                                       & \sqrt{\frac{3}{4\pi}}\, \dlmnarg{1}{-1}{\mz}\,\dlmnarg{1}{-1}{-\spin} & \cdots & \sqrt{\frac{2\elmax+1}{4\pi}}\, \dlmnarg{\elmax}{-1}{\mz}\,\dlmnarg{\elmax}{-1}{-\spin} \\
\sqrt{\frac{1}{4\pi}}\, \dlmnarg{0}{0}{\mz}\,\dlmnarg{0}{0}{-\spin} & \sqrt{\frac{3}{4\pi}}\, \dlmnarg{1}{0}{\mz}\,\dlmnarg{1}{0}{-\spin} & \cdots & \sqrt{\frac{2\elmax+1}{4\pi}}\, \dlmnarg{\elmax}{0}{\mz}\,\dlmnarg{\elmax}{0}{-\spin} \\
                                       & \sqrt{\frac{3}{4\pi}}\, \dlmnarg{1}{1}{\mz}\,\dlmnarg{1}{1}{-\spin} & \cdots & \sqrt{\frac{2\elmax+1}{4\pi}}\, \dlmnarg{\elmax}{1}{\mz}\,\dlmnarg{\elmax}{1}{-\spin} \\
&&\ddots&\vdots\\
                                                                                       &&& \sqrt{\frac{2\elmax+1}{4\pi}}\, \dlmnarg{\elmax}{\elmax}{\mz}\,\dlmnarg{\elmax}{\elmax}{-\spin} \\
\end{pmatrix}
\begin{pmatrix}
 \fst_{0,\mz} \\
  \vdots \\
 \fst_{\elmax,\mz} \\
\end{pmatrix}
\spcend ,
\end{equation*}
where we have used the convention that \dmatsmall-functions with no
argument specified take the assumed argument of $\pi/2$, \linebreak \ie\
\mbox{$\dlms\equiv\dlms(\pi/2)$}, and empty elements of the matrix not
enclosed by ellipsis dots are zero.  In order to solve this equation
for the harmonic coefficients $\shc{\ft}{\el}{\mz}$, we only require
the lower half of the system; the upper half of the system will be
satisfied automatically by symmetry considerations.  We thus consider
only the triangular system
\begin{equation*}
\begin{pmatrix}
 \Fst_{\mz,0} \\
  \vdots \\
 \Fst_{\mz,\elmax} \\
\end{pmatrix}
= \img^{-(\mz+\spin)}
\begin{pmatrix}
\sqrt{\frac{1}{4\pi}}\, \dlmnarg{0}{0}{\mz}\,\dlmnarg{0}{0}{-\spin} & \sqrt{\frac{3}{4\pi}}\, \dlmnarg{1}{0}{\mz}\,\dlmnarg{1}{0}{-\spin} & \cdots & \sqrt{\frac{2\elmax+1}{4\pi}}\, \dlmnarg{\elmax}{0}{\mz}\,\dlmnarg{\elmax}{0}{-\spin} \\
                                       & \sqrt{\frac{3}{4\pi}}\, \dlmnarg{1}{1}{\mz}\,\dlmnarg{1}{1}{-\spin} & \cdots & \sqrt{\frac{2\elmax+1}{4\pi}}\, \dlmnarg{\elmax}{1}{\mz}\,\dlmnarg{\elmax}{1}{-\spin} \\
&&\ddots&\vdots\\
                                                                                       &&& \sqrt{\frac{2\elmax+1}{4\pi}}\, \dlmnarg{\elmax}{\elmax}{\mz}\,\dlmnarg{\elmax}{\elmax}{-\spin} \\
\end{pmatrix}
\begin{pmatrix}
 \fst_{0,\mz} \\
  \vdots \\
 \fst_{\elmax,\mz} \\
\end{pmatrix}
\spcend ,
\end{equation*}
which we write 
\begin{equation}
\label{eqn:matrix_fwd}
{}_\spin \vect{\Ft}_\mz = {}_\spin \mathbf{D}_\mz \: {}_\spin \vect{\ft}_\mz
\spcend ,
\end{equation}
where the $\img^{-(\mz+\spin)}$ factor is incorporated into the
definition of ${}_\spin \mathbf{D}_\mz$.  Once the \Fstfourier\
coefficients are computed, as outlined in
\sectn{\ref{sec:algorithms_fourier_coeff}}, we can then compute
$\sshc{\ft}{\el}{\mz}{\spin}$ for all \el\ by inverting the triangular
system specified by \eqn{\ref{eqn:matrix_fwd}}.
For an arbitrary complex signal \fs, a similar system must be solved
for all \mz.  However, if one is computing for both spin $\pm\spin$ or
for $\spin=0$ for a function \fs\ that satisfies $\fs^\cconj=\fsm$
(\ie\ a real signal for the spin $\spin=0$ case), then it is only necessary
to solve systems for non-negative \mz.  Conjugate symmetry may then be
used to compute the harmonic coefficients for negative \mz.

\ifjnr
\begin{multicols}{2}
\fi

For the spin $\spin=0$ case, additional symmetries exist in the matrix
${}_\spin \mathbf{D}_\mz$ that may be exploited to improve the efficiency of
inversion. 
Recall that $\dlmzhalfpi$ is zero for odd $\el+\m$ (which
follows from \eqn{\ref{eqn:wignerd_sym6}}).  This implies that all
entries of ${}_0 \mathbf{D}_\mz$ are zero for odd $\el+\m\p$.  We
illustrate this point with an example for $\elmax=5$.  In this case
the matrix ${}_0 \mathbf{D}_\mz$ takes the form
\begin{equation*}
{}_0 \mathbf{D}_\mz = 
\begin{pmatrix}
\bullet && \bullet && \bullet \\
& \bullet &&\bullet &&\bullet\\
&&\bullet &&\bullet\\
&&&\bullet &&\bullet\\
&&&&\bullet \\
&&&&&\bullet \\
\end{pmatrix}
\spcend ,
\end{equation*}
where all entries without a dot are zero.  Due to the structure of
${}_0 \mathbf{D}_\mz$ it is possible to separate
\eqn{\ref{eqn:matrix_fwd}} into two independent triangular systems by
extracting out the even and odd rows of the original system.  This
provides a small improvement in the number of computations required to
invert \eqn{\ref{eqn:matrix_fwd}} and is the method employed in our
implementations for the spin $\spin=0$ case.

We have so far discussed how to invert \eqn{\ref{eqn:fourier_coeff}}
for \fst\ on \torus\ in order to recover \fstlm.  Now we briefly
review how to recover the harmonic coefficients \fslm\ of the original
signal \fs\ on \sphere.  Recall that \eqn{\ref{eqn:fourier_coeff}}, and
consequently \eqn{\ref{eqn:matrix_fwd}}, are satisfied by \fste\ and
\fsto\ only for even and odd $\m+\spin$ respectively.  Consequently, the
harmonic coefficients of \fs\ will be given by
\begin{equation}
\fslm=
\begin{cases}
{}_\spin \mathbf{D}_\m^{-1} \: {}_\spin\vect{\Ft}_\m^{\rm e} & \m+\spin \mbox{ even}\\
{}_\spin\mathbf{D}_\m^{-1} \: {}_\spin\vect{\Ft}_\m^{\rm o} & \m+\spin \mbox{ odd}\\
\end{cases} 
\spcend .
\end{equation}

\subsubsection{Inverse transform}

Our fast algorithm may also be used to perform the inverse spherical
harmonic transform, \ie\ to recover a function on the sphere from its
spherical harmonic coefficients \fslm.  In this case we cannot
construct \fstelm\ and \fstolm\ from \fslm\ since we do not know what
values \fstelm\ and \fstolm\ should take for odd and even $\m+\spin$
respectively.  However, this is not problematic since we can
reconstruct an \fst\ that is extended to \torus\ in a different
manner, but which is identical to \fs\ only on the half of \torus\
that maps directly onto \sphere.  It is possible to do this by first
applying \eqn{\ref{eqn:matrix_fwd}} on the harmonic coefficients
\fslm\ directly.  Values for negative $\m\p$ may then be computed
using the symmetry relation specified by \eqn{\ref{eqn:Ffourier_sym}}.
An \fft\ may then be used to recover a sampled version of \fst\
through \eqn{\ref{eqn:idft_spatial}}.  We then recover \fs\ from the
values of \fst\ for which $\saa\in[0,\pi)$.  Note that outside of this
domain \fst\ may not have even or odd symmetry, however we are not
interested in these values and thus discard them.

\section{Complexity analysis and numerical experiments}
\label{sec:experiments}

The fast forward and inverse spin spherical harmonic transforms
derived in the previous section have been implemented in double
precision arithmetic using the
FFTW\footnote{\url{http://www.fftw.org/}} library to compute Fourier
transforms.  We discuss here the theoretical memory and computational
requirements of these algorithms, before performing numerical
experiments to evaluate execution time and numerical stability.  To
avoid unnecessary complexity we restrict the numerical experiments to
real spin $\spin=0$ functions on the sphere.  Due to the structure of
our algorithms (in particular, since the spin number \spin\ is simply a
parameter of our algorithms), memory requirements and computational cost
for any spin value will be identical to the spin $\spin=0$ case, while
numerical stability will be similar.

\subsection{Memory and computational complexity}
\label{sec:experiments_complexity}

It is possible to reduce the computational burden of our fast
transforms by precomputing the reduced Wigner \dmatsmall-functions for
an argument of $\pi/2$ over a range of $\el$, $\m$ and $\n$ values.
This type of precomputation compares favourably with the precomputation
required for other fast scalar spherical harmonic transforms, which
require a precomputation of associated Legendre functions for a range
of $\el$ and $\m$ indices, \emph{and} over the $\saa$ values of a
discretised grid.  Our precomputation of the \dmatsmall-functions is
independent of the grid: as the band-limit $\elmax$ increases we must
precompute more terms, but it is not necessary to re-evaluate the
precomputed terms already calculated for a lower band-limit.  This is
not the case when precomputing associated Legendre functions over the
\saa\ grid.
When evaluating the \dmatsmall-functions it is only necessary to
precompute $\dlmnhalfpi$ for non-negative $\m$ and $\n$.  Values for
negative $\m$ and $\n$ indices may then be recovered by noting the
symmetry relations \eqn{\ref{eqn:wignerd_sym1}} and
\eqn{\ref{eqn:wignerd_sym6}}.  Naively $\elmax^3$ terms must be
computed, however this ignores the restriction that $|\m| \leq \el$
and $|\n| \leq \el$ for each \el.  Taking this restriction into
consideration the number of precomputed terms is
\begin{equation*}
\sum_{\el=0}^{\elmax} \: (\el+1)^2 = \elmax^3/3 + 3\elmax^2/2 + 13 \elmax/6
+ 1
\spcend .
\end{equation*}
To precompute all of the associated Legendre terms (for $\m \geq 0$
only and noting the restriction that $|\m| \leq \el$ for each \el)
requires the computation of $\elmax^3$ terms, when \saa\ is defined
over a grid of size $2\elmax$.  Although the complexity of both
precomputations scale as $\order(\elmax^3)$, precomputing the reduced
Wigner \dmatsmall-functions requires approximately one third of the
number of terms as precomputing the associated Legendre functions.

The overall memory requirements of our fast spin spherical harmonic
transform algorithms are dominated by the storage of the precomputed
reduced Wigner \dmatsmall-functions and therefore scale as
$\order(\elmax^3)$.  These precomputed data could be read from disk
rather than stored in memory, however this would reduce the speed of
the algorithms considerably.  In this setting the memory requirements
scale as $\order(\elmax^2)$.

We now consider the computational complexity of our algorithms.
Although we have discussed the details of the algorithms in
\sectn{\ref{sec:algorithms_fwdinv}}, the algorithms essentially equate
to directly evaluating or inverting \eqn{\ref{eqn:sht_inv_fast}} and
\eqn{\ref{eqn:fourier_coeff}} as derived in
\sectn{\ref{sec:algorithms_harmonic}}.  All other operations do not
alter the overall complexity of the algorithms and merely introduce a
prefactor.  By applying an \fft\ it is possible to reduce the
complexity of evaluating/inverting \eqn{\ref{eqn:sht_inv_fast}}
from $\order(\elmax^4)$ to $\order(\elmax^2 \log_2 \elmax)$.  The
linear system defined by \eqn{\ref{eqn:fourier_coeff}} involves a
triangular matrix, as specified explicitly by
\eqn{\ref{eqn:matrix_fwd}}.  A triangular system of size $\elmax
\times \elmax$ may be evaluated/inverted in
$\order(\elmax^2)$ operations.  This system must be considered for
each value of $\mz$, hence the overall complexity of evaluating or
inverting \eqn{\ref{eqn:fourier_coeff}} is $\order(\elmax^3)$.  The
complexity of both of the forward and inverse fast spin spherical
harmonic transforms is thus
\ifjnr
\begin{equation*}
\underbrace{\order(\elmax^2 \log_2 \elmax)}_{\mbox{\scriptsize Evaluation/inversion of
  \eqn{\ref{eqn:sht_inv_fast}} by \fft}} 
\quad+\quad 
\underbrace{\quad\order(\elmax^3)\quad}_{\mbox{\scriptsize Evaluation/inversion of \eqn{\ref{eqn:fourier_coeff}}}}
\spcend ,
\end{equation*}
resulting in an overall complexity of $\order(\elmax^3)$.
\else
\begin{equation*}
\underbrace{\order(\elmax^2 \log_2 \elmax)}_{\mbox{\scriptsize Evaluation/inversion of
  \eqn{\ref{eqn:sht_inv_fast}} by \fft}} 
\quad+\quad 
\underbrace{\quad\order(\elmax^3)\quad}_{\mbox{\scriptsize Evaluation/inversion of \eqn{\ref{eqn:fourier_coeff}}}}
\quad\quad \sim \quad\quad
\underbrace{\quad\order(\elmax^3)\quad}_{\mbox{\scriptsize Overall
    complexity}}
\spcend .
\end{equation*}
\fi
Using our implementation of these algorithms with precomputed data
stored in memory, we evaluate computation times for real spin $\spin=0$
signals over a range of band-limits.  These timing tests are performed
on random band-limited test functions on the sphere.  The test
functions are defined through their spherical harmonic coefficients,
with independent real and imaginary parts distributed uniformly over
the interval $[-1,1]$.  Six band-limits from $\elmax=16$ to
$\elmax=512$ are considered, increasing by a factor of two at each
step.  Computation time measurements are performed on a laptop with a
2.2GHz Intel Core Duo processor and 2GB of RAM and are averaged over
five random test functions.  \fig{\ref{fig:timing}} shows a 
plot of the average computation time in seconds $\tau$ with band-limit.  Also
shown on \fig{\ref{fig:timing}} is the slope corresponding to
computation time that scales as $\elmax^3$.  The average computation
time for the implementation of our fast spin spherical harmonic
transforms scales approximately as $\elmax^3$, as predicted by the
theoretical computational complexity discussed above.

\ifjnr
\vspace*{8mm}
\begin{figurehere}
\else
\begin{figure}[t]
\fi
\centering
\includegraphics[width=80mm]{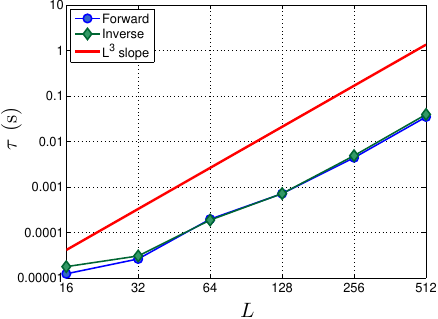}
\caption{Average computation time $\tau$ (in seconds) for the forward and inverse
  transforms of random real spin $\spin=0$ test functions.
  Note that the computation time scales like $\elmax^3$ as expected.
 }
\label{fig:timing}
\ifjnr
\end{figurehere}
\else
\end{figure}
\fi

\subsection{Numerical stability}

The numerical stability of our algorithms is examined in this section.
Unfortunately the algorithms are found to be stable only for very low
band-limits in the order of $\elmax \simeq 32$ and below.  We first
evaluate the stability of the algorithms in detail, before examining
the source of the instability and possible solutions.

To test the stability of our algorithms we perform the following
numerical tests, restricted here to real spin $\spin=0$ functions on
the sphere.  Firstly, we consider a random test function on the sphere
defined through its harmonic coefficients, with independent real and
imaginary parts distributed uniformly over the interval $[-1,1]$ (the
same test functions considered in
\sectn{\ref{sec:experiments_complexity}}).  By defining the test
function explicitly in harmonic space we ensure that it is
band-limited.  We then perform the inverse spherical harmonic
transform to compute the associated real values of the function on the
sphere, before performing the forward spherical harmonic transform to
reconstruct the spherical harmonic coefficients of the function.  The
error between the original spherical harmonic coefficients
$\shc{f}{\el}{\m}$ and the reconstructed spherical harmonic
coefficients $\shc{g}{\el}{\m}$ is examined to assess the numerical
stability of the algorithms.
Five different error metrics are computed, including the maximum
absolute error
\begin{equation}
\epsilon_{\rm max} = \mathop{\rm max}_{\el,\m} \:
\bigl | \shc{f}{\el}{\m} - \shc{g}{\el}{\m} \bigr |
\spcend ,
\end{equation}
the mean absolute error
\begin{equation}
\epsilon_{\rm mean} = \mathop{\rm mean}_{\el,\m} \:
\bigl | \shc{f}{\el}{\m} - \shc{g}{\el}{\m} \bigr |
\spcend ,
\end{equation}
the median absolute error
\begin{equation}
\epsilon_{\rm median} = \mathop{\rm median}_{\el,\m} \:
\bigl | \shc{f}{\el}{\m} - \shc{g}{\el}{\m} \bigr |
\spcend ,
\end{equation}
the root-mean-square error
\begin{equation}
\epsilon_{\rm rms} = \Bigl[ \mathop{\rm mean}_{\el,\m} \:
\bigl | \shc{f}{\el}{\m} - \shc{g}{\el}{\m} \bigr |^2 \Bigr]^{1/2}
\end{equation}
and the relative root-mean-square error
\begin{equation}
\epsilon_{\rm rel \: rms} = 
\Bigl[
\mathop{\sum}_{\el,\m} \:
\bigl | \shc{f}{\el}{\m} - \shc{g}{\el}{\m} \bigr |^2
\:
\bigl /
\:
\mathop{\sum}_{\el,\m} \:
\bigl | \shc{f}{\el}{\m} \bigr |^2 \Bigr]^{1/2}
\spcend .
\end{equation}
Error measurements for seven band-limits from $\elmax=8$ to
$\elmax=512$, increasing by a factor of two at each step, averaged
over five random test functions on the sphere are illustrated in
\fig{\ref{fig:errors}}.  For very low band-limits the errors are on
the order of the machine precision, however as the band-limit
increases the instability of the algorithms become apparent as the
errors quickly increase to unacceptable levels.  From the median error
it is apparent that for band-limits below $\elmax\simeq64$ many of the
spherical harmonic coefficients are well recovered and only a
relatively small number of coefficients are responsible for the large
errors measured by the other error metrics.  However, above this
band-limit even the median error increases to an unacceptable level
indicating that most coefficients are poorly recovered.
\begin{figure*}[t]
\centering
\mbox{
\subfigure[All band-limits]{\includegraphics[width=80mm]{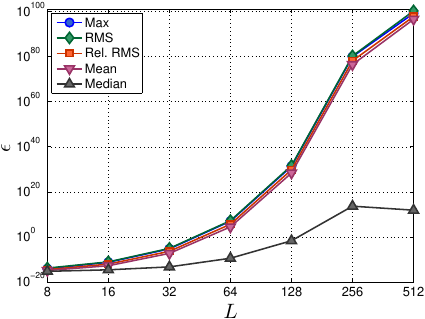}}
\quad \quad 
\subfigure[Lower band-limits]{\includegraphics[width=80mm]{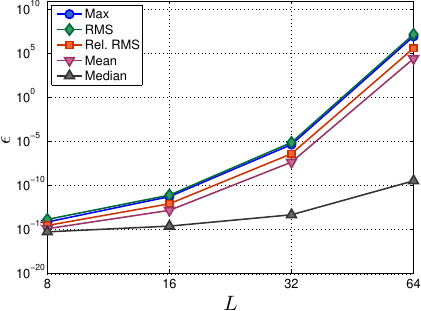}}
}
\caption{Average errors, for various error metrics (defined in the text), between the original and reconstructed
  spherical harmonic coefficients of random real spin $\spin=0$
  functions for the numerical stability experiment explained in the
  text.}
\label{fig:errors}
\end{figure*}
To examine the location of the harmonic coefficients that are poorly
recovered we plot in \fig{\ref{fig:harmonic_coefficients}} the
absolute value of the original harmonic coefficients $|\shc{f}{\el}{\m} |$ and the
residual error $|\shc{g}{\el}{\m} - \shc{f}{\el}{\m}|$ for one test
function for various band-limits.  For $\elmax=16$ and $\elmax=32$
(and $\elmax=8$ although not shown) all harmonic coefficients are
recovered accurately.  For $\elmax=64$ many coefficients are recovered
accurately, however a small number of coefficients near the diagonal
are not.  Most coefficients are recovered poorly for band-limits above
$\elmax\simeq128$.  Errors are first introduced along the diagonal
where $\el\sim\m$ and then quickly spread as $\el$ increases.

\newlength{\matrixwidth}
\setlength{\matrixwidth}{38mm}

\newlength{\matrixwidthtwo}
\setlength{\matrixwidthtwo}{39.5mm}

\begin{figure*}[t]
\begin{minipage}{\textwidth}
\begin{multicols}{2}

\centering
\mbox{
\subfigure[Original $\elmax=16$]{\includegraphics[width=\matrixwidth]{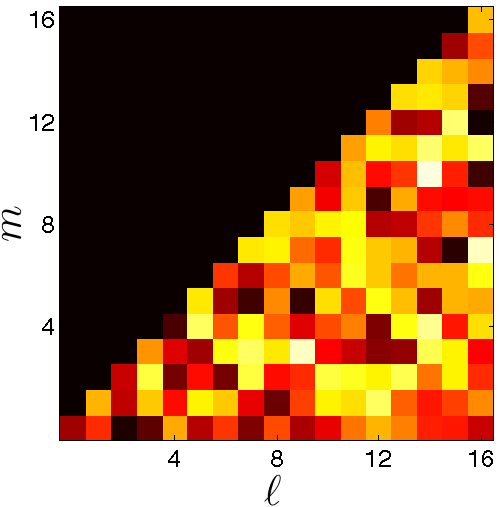}}
\quad
\subfigure[Residual $\elmax=16$]{\includegraphics[width=\matrixwidth]{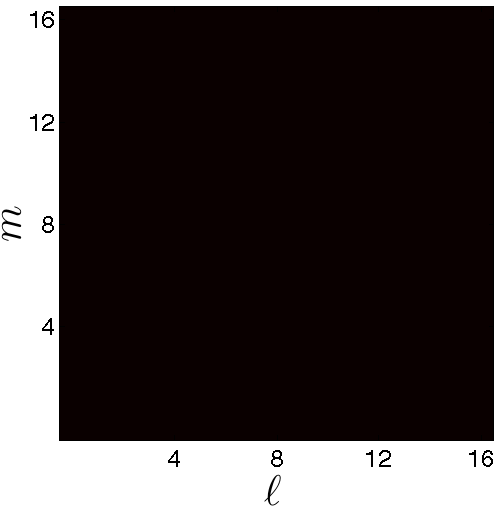}}
} \\

\mbox{
\subfigure[Original $\elmax=32$]{\includegraphics[width=\matrixwidth]{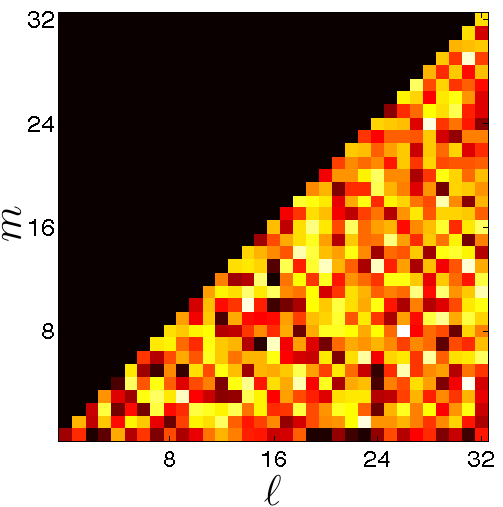}}
\quad
\subfigure[Residual $\elmax=32$]{\includegraphics[width=\matrixwidth]{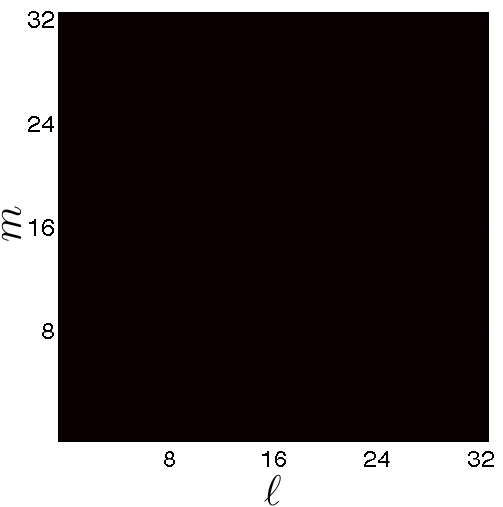}}
} \\

\mbox{
\subfigure[Original $\elmax=64$]{\includegraphics[width=\matrixwidth]{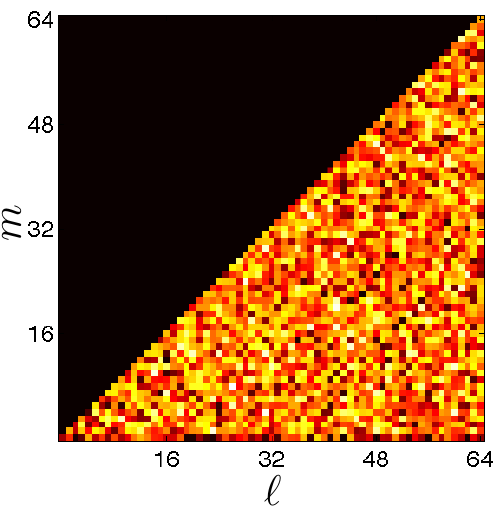}}
\quad
\subfigure[Residual $\elmax=64$]{\includegraphics[width=\matrixwidth]{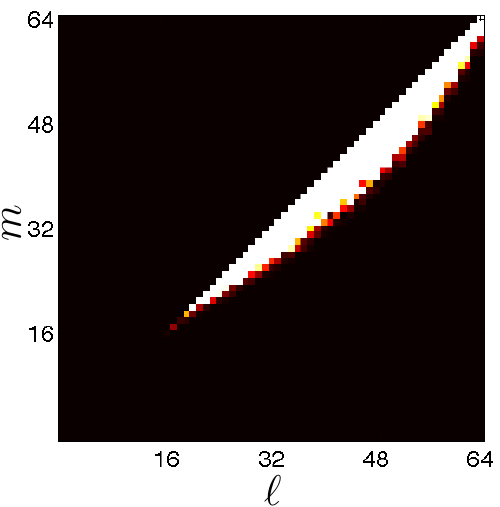}}
} \\

\mbox{
\subfigure[Original $\elmax=128$]{\includegraphics[width=\matrixwidthtwo]{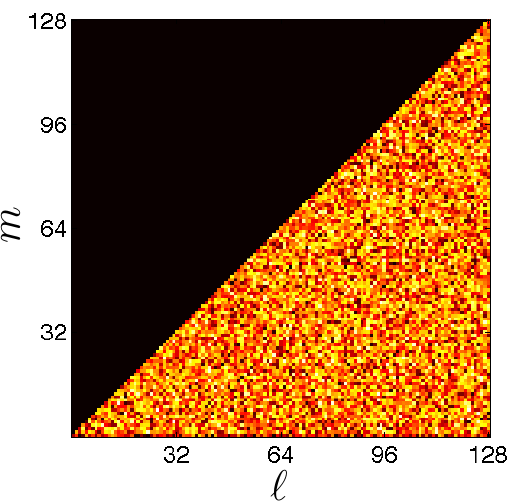}}
\quad
\subfigure[Residual $\elmax=128$]{\includegraphics[width=\matrixwidthtwo]{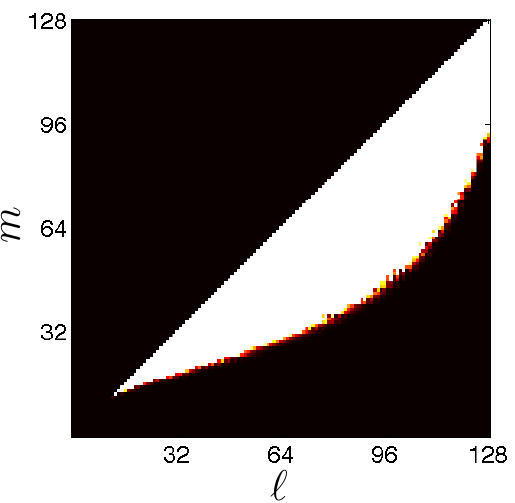}}
} \\

\mbox{
\subfigure[Original $\elmax=256$]{\includegraphics[width=\matrixwidthtwo]{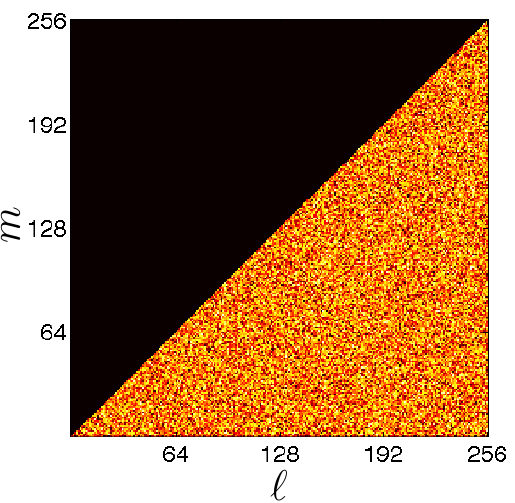}}
\quad
\subfigure[Residual $\elmax=256$]{\includegraphics[width=\matrixwidthtwo]{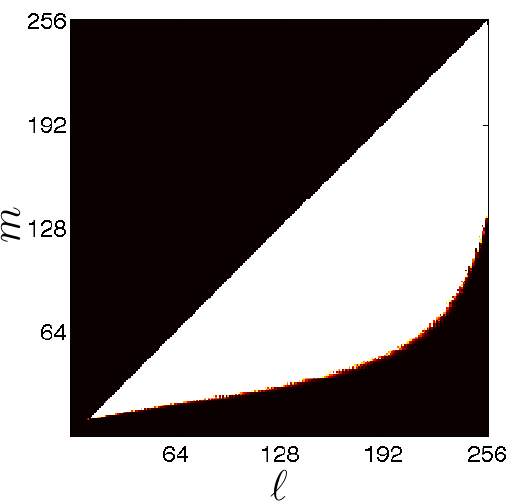}}
} \\

\mbox{
\subfigure[Original $\elmax=512$]{\includegraphics[width=\matrixwidthtwo]{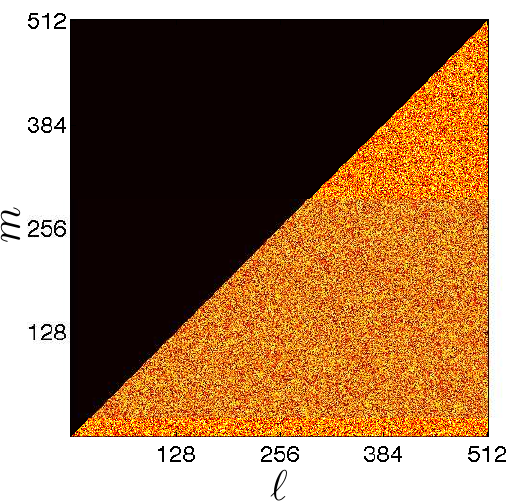}}
\quad
\subfigure[Residual $\elmax=512$]{\includegraphics[width=\matrixwidthtwo]{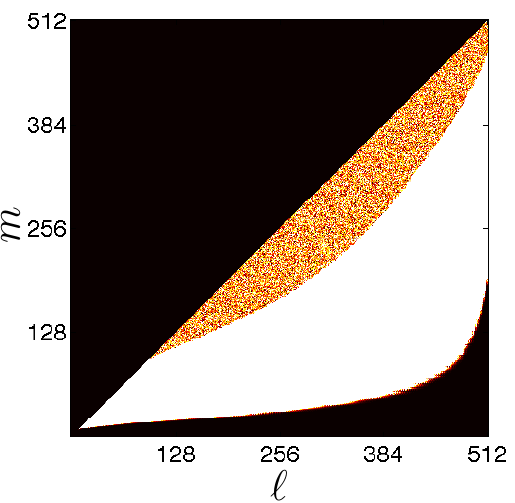}}
} 

\end{multicols}
\end{minipage}
\caption{Absolute value of original spherical
  harmonic coefficients and residual errors for the numerical experiments explained in the
  text.  The same colour scale is used for all plots, ranging over $[0,\sqrt{2}]$, with zero
  represented by dark red/black and values greater or equal to $\sqrt{2}$ by white, with
  intermediate values interpolated between these extremes.}
\label{fig:harmonic_coefficients}
\end{figure*}

The source of the numerical stability seen above has been traced to
the system defined by \eqn{\ref{eqn:fourier_coeff}}, which is
specified explicitly as a matrix system in \eqn{\ref{eqn:matrix_fwd}}.
As an example, in \fig{\ref{fig:illmat}} we show the triangular matrix
${}_0 \mathbf{D}_\mz$ defining this system for $\elmax=50$ and
$\mz=38$.  The matrix is extremely poorly conditioned with condition
number $6\times10^{14}$.
When solving these triangular systems the increasingly large off-diagonal
terms cause numerical errors to explode when recovering harmonic coefficients
for low \el, resulting in the error structure exhibited in
\fig{\ref{fig:harmonic_coefficients}}. 

We have investigated a number of numerical methods for solving these
ill-conditioned systems, including a singular value decomposition
(SVD) \cite{press:1992}, preconditioning \cite{press:1992}, the generalised minimal residual method
(GMRES) \cite{press:1992}, iterative refinements \cite{press:1992} and various re-normalisations; however,
without success.  It may indeed be possible to solve these systems
using alternative numerical methods, hence we
make data defining the system shown in \fig{\ref{fig:illmat}}, with the correct ${}_0
\vect{\Ft}_\mz$ and ${}_0 \vect{\ft}_\mz$ vectors, available for
download\footnote{\url{http:www.mrao.cam.ac.uk/~jdm57/research/fsht.html}}
in case others wish to examine the system and have more success in
solving it.  Note that, although we have examined the numerical
stability of the spin $\spin=0$ case only, our algorithms will also be
unstable for other spin values since the corresponding linear systems
that must be inverted will exhibit similar structure to the system
displayed in \fig{\ref{fig:illmat}} (\ie\ increasingly large
off-diagonal terms).
It may be possible to rewrite the original
formulation of our fast algorithms so that the equations are
normalised in such a way as to avoid these ill-conditioned matrices,
however we have as yet been unable to do so.  Nevertheless, the
conditioning of these systems is only a problem when inverting the
system, thus it is only the fast forward spin spherical harmonic
transform that suffers from this instability problem.  The inverse
spin spherical harmonic transform is likely to be accurate to high
precision up to band-limits where the reduced Wigner
\dmatsmall-functions can be computed accurately.  However, we cannot
test the inverse transform in isolation since a stable forward
transform that is exact does not exist
currently on the $(\saa_\saai, \sab_\sabi)$ grid positions defined by
\eqn{\ref{eqn:nodes_theta}} and \eqn{\ref{eqn:nodes_phi}}.


\ifjnr
  \vspace*{1.2mm}
  \begin{figurehere}
\else
  \begin{figure*}[t]
\fi
\centering
\ifjnr
\includegraphics[width=68mm]{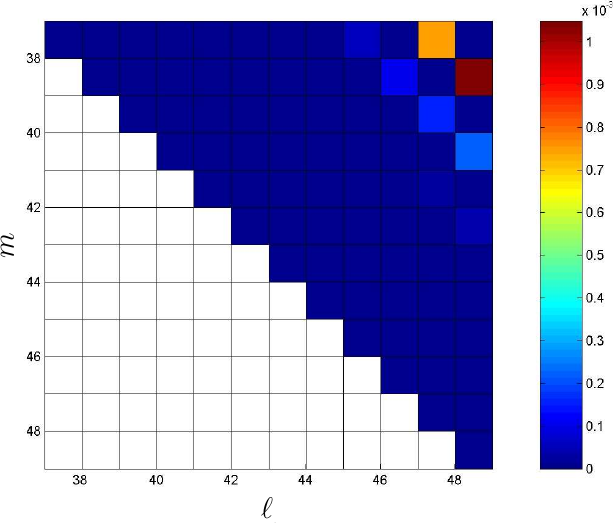}
\vspace*{-3mm}
\else
\includegraphics[width=68mm]{figures/illmat_bar_psfrag}
\fi
\caption{Absolute values of the ill-conditioned triangular matrix of
  the system that must be solved for $\elmax=50$ and $\mz=38$.  The
  matrix is extremely poorly conditioned with condition number
  $6\times10^{14}$.  }
\label{fig:illmat}
\ifjnr
  \end{figurehere}
\else
  \end{figure*}
\fi

\section{Conclusions}
\label{sec:conclusions}

Algorithms have been derived to perform the forward and inverse
spherical harmonic transform of functions on the sphere with arbitrary
spin number. 
These algorithms involve recasting the spin transform on the sphere as
a Fourier transform on the torus.  \fft s are then
used to compute Fourier coefficients, which are related to spherical
harmonic coefficients through a linear transform.  By recasting the
problem as a Fourier transform on the torus we appeal to the usual
Shannon sampling theorem to develop spherical harmonic transforms that
are theoretically exact for band-limited functions, thereby providing
an alternative sampling theorem on the sphere.

Previous approaches to performing spin transforms on the sphere have
appealed to the spin lowering and raising operators to relate spin
transforms to scalar transforms.  For spin $\spin$ transforms based on
this method the lowest computational complexity that can be attained
scales as $\order(\spin \elmax^2 \log_2^2 \elmax)$, whereas our
algorithm scales as $\order(\elmax^3)$ for arbitrary spin number
\spin.
Moreover, our algorithm is general in the sense that the spin number
is merely a parameter of the algorithm; altering the spin number does
not alter the structure of the algorithm.  Furthermore, the algorithms
apply for arbitrary band-limit \elmax, whereas many other methods are
restricted to band-limits that are a power of two.

Our algorithms have been implemented.  The computational speed of the
implementation may be improved in future by using fast discrete cosine
and sine transforms in place of \fft s, although this will not alter the
scaling of computation time with band-limit.  Numerical tests showed
that computation time scales as $\order(\elmax^3)$, as expected.
Numerical tests were also performed to assess the stability of the
algorithms.  Unfortunately the algorithms were found to be unstable
for band-limits above $\elmax\simeq32$.  The source of the instability
was determined to be due to the poor conditioning of the linear system
relating the Fourier and spherical harmonic coefficients.
Consequently, we expect only the forward transform to be unstable and
not the inverse transform, however it is not possible to verify this
hypothesis since a stable forward transform that is exact (to
numerical precision) does not exist currently on our pixelisation of
the sphere.  A number of numerical methods have been applied in an
attempt to solve the ill-conditioned system but these have not proved
successful.  


A fast algorithm to perform a scalar spherical harmonic transform has
been developed by Anthony Lasenby (private communication) in the
context of geometric algebra.  This algorithm shares many of the
properties of the algorithms derived in this work (including,
unfortunately, hints of numerical stability issues) and may simply be
an alternative formulation, for the spin $\spin=0$ case, of the algorithms
derived herein.  If this indeed proves to be the case, then this
alternative geometric algebra formulation may provide insight on
alternative approaches to reformulate our algorithms to improve
numerical stability.
In ongoing work we are investigating the numerical stability problem
in more detail and are attempting to renormalise or reformulate the
algorithms in such a way as to eliminate this instability.





\ifjnr
  \bibliographystyle{jnr}
\else
  \bibliographystyle{habbrv}
\fi
\setlength{\bibsep}{1mm}
\bibliography{bib}






\ifjnr
  \end{multicols}
\fi
\end{document}